\documentclass[11pt]{article}
\usepackage{times,amsfonts,latexsym}

\catcode`\@=11
\def\marginnote#1{}

\newcount\hour
\newcount\minute
\newtoks\amorpm
\hour=\time\divide\hour by60
\minute=\time{\multiply\hour by60 \global\advance\minute by-\hour}
\edef\standardtime{{\ifnum\hour<12 \global\amorpm={am}%
        \else\global\amorpm={pm}\advance\hour by-12 \fi
        \ifnum\hour=0 \hour=12 \fi
      \number\hour:\ifnum\minute<10 0\fi\number\minute\the\amorpm}}
\edef\militarytime{\number\hour:\ifnum\minute<10 0\fi\number\minute}

\def\draftlabel#1{{\@bsphack\if@filesw {\let\thepage\relax
   \xdef\@gtempa{\write\@auxout{\string
      \newlabel{#1}{{\@currentlabel}{\thepage}}}}}\@gtempa
   \if@nobreak \ifvmode\nobreak\fi\fi\fi\@esphack}
        \gdef\@eqnlabel{#1}}
\def\@eqnlabel{}
\def\@vacuum{}
\def\draftmarginnote#1{\marginpar{\raggedright\scriptsize\tt#1}}

\def\draft{\oddsidemargin -0.1truein
        \def\@oddfoot{\sl preliminary draft \hfil
        \rm\thepage\hfil\sl\today\quad\militarytime}
        \let\@evenfoot\@oddfoot \overfullrule 3pt
        \let\label=\draftlabel
        \let\marginnote=\draftmarginnote
\def\@eqnnum{{\rm (\theequation)}
\rlap{\kern\marginparsep\tt\@eqnlabel}%
\global\let\@eqnlabel\@vacuum}  }
\def\underset#1#2{\mathrel{\mathop{#2}\limits_{#1}}}
\def\overset{\stackrel}
 
\makeatletter
\@addtoreset{equation}{section}
\makeatother
\renewcommand{\theequation}{\thesection.\arabic{equation}}

\newfont{\blackb}{msbm10 scaled\magstep1}
 
\newcounter{subequation}[equation]
\makeatletter

\expandafter\let\expandafter
\reset@font\csname reset@font\endcsname

\def\subeqnarray{\arraycolsep1pt
    \def\@eqnnum\stepcounter##1{\stepcounter{subequation}%
        {\reset@font\rm(\theequation\alph{subequation})}}
\jot5mm     \eqnarray}

\def\subarray{\arraycolsep1pt
    \def\@eqnnum\stepcounter##1{\stepcounter{subequation}%
        {\reset@font\rm(\alph{subequation})}}
\jot5mm     \eqnarray}

\makeatother

\newfont{\calig}{cmsy10 scaled\magstep1}

\def\text#1{\hbox{#1}}
 
\newtheorem{theorem}{Theorem}[section]

\newtheorem{remark}{Remark}[section]
\newtheorem{corollary}{Corollary}[section]
\newtheorem{lemma}{Lemma}[section]
\newtheorem{definition}{Definition}[section]
 
\newcommand{\qed}{\hfill$\Box$}

\newcommand{\be}{\begin{equation}}
\newcommand{\ee}{\end{equation}}
\newcommand{\ben}{\begin{displaymath}}
\newcommand{\een}{\end{displaymath}}
\newcommand{\ba}{\begin{array}}
\newcommand{\ea}{\end{array}}
\newcommand{\baa}{\begin{eqnarray}}
\newcommand{\eaa}{\end{eqnarray}}
\newcommand{\baan}{\begin{eqnarray*}}
\newcommand{\eaan}{\end{eqnarray*}}

\newcommand{\non}{\nonumber\\}

\newcommand{\mathon}{\mathversion{bold}}
\newcommand{\mathoff}{\mathversion{normal}}
\def\a{\alpha}
\def\b{\beta}

\def\D{\Delta}
\def\d{\delta}
\def\e{\varepsilon}

\def\ka{\kappa}
\def\l{\lambda}
\def\L{\Lambda}

\def\ph{\phi}
\def\r{\rho}
\def\s{\sigma}

\def\Th{\Theta}
\def\o{\omega}
\def\Om{\Omega}

\def\half{\textstyle{\frac12}}
\def\ie{{\it i.e.}}

\def\la{\label}
\def\Ref{\ref}
\def\c{\cite}
\def\f{\frac}
\def\p{\partial}
\def\Ref#1{(\ref{#1})}

\def\res{{\mathrm res}}
\def\tPsi{{\widetilde{\Psi}}}
\def\tpsi{{\widetilde{\psi}}}

\def\str{{\rm str}}

\def\hb{{\widehat{\b}}}
\def\vm{{\vec{\mu}}}
\def\vz{{\vec{z}}}
\def\vph{\varphi}
\def\Ka{{\mathcal{K}}}
\def\wtt {{\widetilde{t}}}
\def\tPsi{{\widetilde{\Psi}}}
\begin{document}

\begin{titlepage}

\begin{center}

{\bf \huge Trigonometric $osp(1|2)$ Gaudin model}

\bigskip

{\sc P. ~P.~Kulish
\footnote{E-mail address: kulish@pdmi.ras.ru \,;
On leave of absence from Steklov Mathematical Institute, 
Fontanka 27, 191011, St.Petersburg, Russia.} 
and N. ~Manojlovi\'c 
\footnote{E-mail address: nmanoj@ualg.pt}\\ 
\medskip
{\it \'Area Departamental de Matem\'atica, F. C. T., 
Universidade do Algarve\\ Campus de Gambelas, 8000-117 Faro, 
Portugal}}\\ 

\vskip2.5cm 

\end{center}

\begin{abstract}

The problems connected with Gaudin models are reviewed
by analyzing model related to the trigonometric  $osp(1|2)$
classical $r$-matrix. The eigenvectors of the trigonometric 
$osp(1|2)$ Gaudin Hamiltonians are found using explicitly 
constructed creation operators. The commutation relations 
between the creation operators and the generators of the trigonometric 
loop superalgebra are calculated. The coordinate representation of 
the Bethe states is presented. The relation between the 
Bethe vectors and solutions to the Knizhnik-Zamolodchikov 
equation yields the norm of the eigenvectors. The generalized 
Knizhnik-Zamolodchikov system is discussed both in the rational 
and in the trigonometric case.
\end{abstract} 

\end{titlepage}

\clearpage \newpage



\section{Introduction} 

Classifying integrable systems solvable in the framework of the quantum 
inverse scattering method \c{F95,KS, KBI} by underlying dynamical symmetry 
algebras, one could say that the Gaudin\break  models are the simplest ones being 
based on loop algebras and classical $r$-matrices. More\break  sophisticated 
solvable models correspond to more complicated algebras: Yangians, 
quantum affine algebras, elliptic quantum groups, dynamical quantum groups, etc.  

Gaudin models \c{G,GB} are related to classical $r$-matrices, 
and the density of Gaudin\break 
Hamiltonians 
\be \la{Gh}  
H^{(a)} = \sum_{b \neq a} ^N r_{ab}(z_a - z_b) 
\ee 
coincides with the $r$-matrix. Condition of their commutativity 
$[H^{(a)} , H^{(b)} ] = 0$
is nothing else but the classical Yang-Baxter equation (YBE)
\be \la{cYB}
[r_{ab}(z_a - z_b) , r_{ac}(z_a - z_c) + r_{bc}(z_b - z_c)] + 
[r_{ac}(z_a - z_c) , r_{bc}(z_b - z_c)] = 0 \;,
\ee
where $r$ is antisymmetric and belongs to the tensor product 
$\mathfrak{g} \otimes \mathfrak{g}$ of a Lie algebra $\mathfrak{g}$, 
or its representations and the indices fix the corresponding factors 
in the $N$-fold tensor product of this algebra (see Section 2).

The Gaudin models (GM) related to classical $r$-matrices of simple Lie algebras 
were studied intensively (see \c{GB,S87,S99,J,HKW,B,BF,FFR,ReshVar,ST} and 
references therein). The spectrum and eigenfunctions were found using different 
methods (coordinate and algebraic Bethe Ansatz \c{GB,S87}, separated variables
\c{S87}, etc.). The correlation functions were evaluated for $\mathfrak{g} = sl(2)$
by the Gauss factorization approach \c{S99}. A relation to the Knizhnik-Zamolodchikov 
(KZ) equation of conformal filed theory was established \c{BF,FFR,ReshVar}.

There exists a variety of classical  $r$-matrices with trigonometric
dependence on spectral\break parameter. Although algebraic construction of
integrals of motion is straightforward the\break calculation of the spectrum
and the corresponding eigenstates, by means of algebraic Bethe Ansatz,
still depends on the underlining Lie algebra and $r$-matrix. Moreover, 
many trigonometric $r$-matrices are invariant under diagonal action 
of the Cartan subalgebra only
\be \la{inv}
\left[ h_a + h_b \; , \; r_{ab} (z_a - z_b) \right] = 0 \,,
\ee
as opposed to the rational case where classical $r$-matrix is invariant
under the action of the whole Lie algebra $\mathfrak{g}$. Hence, one can 
modify the Gaudin Hamiltonian \Ref{Gh} by adding a local generator of 
the Cartan subalgebra
\be \la{mod}
H_a \to \widetilde{H}^{(a)} = g \, h _a +  H^{(a)} \,.  
\ee
This modification does not change the creation operators, but the Bethe 
equations and solutions to the KZ system. However, the dependence on the
parameter $g$ (a magnetic field) will be described by a difference 
equation \c{TV2,EV}.

The aim of this paper is to review problems connected with Gaudin models
by analyzing the model related to the trigonometric  $osp(1|2)$ classical
$r$-matrix. Results obtained here are in many respect similar to the
ones we obtained in the case of $osp(1|2)$-invariant rational $r$-matrix \c{KM}.
However, connection of Gaudin model with magnetic field and KZ equations 
requires modification of the later by adding dynamical difference equation 
\c{TV2,EV}.

There are additional peculiarities of Gaudin models related to classical 
$r$-matrices based on Lie superalgebras due to $Z_2$-grading of representation 
spaces and operators. The study of the $osp(1|2)$-invariant Gaudin model 
corresponding to the simplest non-trivial super-case of the $osp(1|2)$ 
invariant $r$-matrix \c{K85} started in \c{BM}. The spectrum of the 
$osp(1|2)$ invariant Gaudin Hamiltonians $H^{(a)}$  was given, antisymmetry 
property of their eigenstates was claimed, and a two site model was connected 
with some physically interesting one (a Dicke model). Let us also
point out that recently rational and trigonometric $sl(2)$ Gaudin models
were used to describe different physical phenomena in metallic grains
\c{ALO} and a condensate fragmentation of confined bosons \c{DS}.
Connection with perturbed WZNW models of conformal field theory was
found in \c{S}.

The creation operators used in the $sl(2)$ Gaudin model (and similarly for
$sl(n)$ case) coincide with one of the $L$-matrix entry \c{GB,S87}. 
However, in the \(osp(1|2)\) case, as it was shown for rational 
$r$-matrix \c{KM}, the creation operators are complicated 
polynomials of the two generators $X^+(\l)$ and 
$v^+(\mu)$ of the loop superalgebra. We introduce $B$-operators 
belonging to the Borel subalgebra of the trigonometric loop superalgebra 
$\mathcal {L}_t (osp(1|2))$ by a recurrence relation. 
Acting on the lowest spin vector (bare vacuum) $B_M(\mu _1 , \dots , \mu_M) \Om _-$
the $B$-operators generate exact eigenstates of the Gaudin Hamiltonians 
$H^{(a)}$, provided Bethe equations are imposed on parameters 
$\{ \mu_j \}$ of the states.
For this reason the $B$-operators are sometimes refereed to as the
creation operators and the eigenstates as the Bethe vectors, or
simply $B$-vectors. Furthermore, the recurrence relation is solved 
explicitly and the commutation relations between the 
$B$-operators and the generators of the loop superalgebra $\mathcal {L}_t (osp(1|2))$ 
as well as the generators of the global superalgebra $osp(1|2) \subset 
\mathcal {L}_t (osp(1|2))$ are calculated. We prove that the constructed 
states are eigenvectors of the generator of the global Cartan subalgebra
$h _{gl}$, but the $B$-vectors are not the lowest spin vectors anymore,
as it was the case for the invariant model \c{TF, KM}. Analogously to the 
rational case \c{KM}, a striking coincidence between the spectrum of the 
$osp(1|2)$ invariant Gaudin Hamiltonians of spin $s$ and the spectrum of 
the Hamiltonians of the $sl(2)$ Gaudin model of the integer spin $2s$ is 
also confirmed in the trigonometric case.

A connection between the $B$-states, when the Bethe equations are not imposed 
on their parameters (``off-shell Bethe states''), of the Gaudin models for simple 
Lie algebras to the solutions of the Knizhnik-Zamolodchikov equation was established 
in the papers \c{BF, FFR}. An explanation of this connection based on Wakimoto modules 
at critical level of the underlying affine algebra was given in \c{FFR}. An explicit 
form of the Bethe vectors in the coordinate representation was given in both papers 
\c{BF, FFR}. The coordinate Bethe Ansatz for the $B$-states of the $osp(1|2)$ Gaudin 
model is obtained in our paper as well. Using commutation relations between the 
$B$-operators and the transfer matrix $t(\l)$, as well as the Hamiltonians 
$H ^{(a)}$, we give an algebraic proof of the fact that explicitly constructed 
$B$-states yield a solution to the Knizhnik-Zamolodchikov equation corresponding 
to a conformal field theory. This connection permits us to calculate 
the norm of the eigenstates of the Gaudin Hamiltonians.
An analogous connection is expected between quantum $osp(1|2)$ spin 
system related to the graded Yang-Baxter equation \c{K85,M,LS,STsu} and quantum 
Knizhnik-Zamolodchikov equation following the lines of \c{TV}. 
We point out possible modifications of the Gaudin Hamiltonians 
and corresponding modifications of the Knizhnik-Zamolodchikov equation, 
similar to the case of the $sl(2)$ Gaudin model which was interpreted in 
\c{R92,BK} as a quantization of the Schlesinger system for isomonodromy
deformation.

The norm and correlation functions of the $sl(2)$ invariant Gaudin model 
were evaluated in \c{S99} using Gauss factorization of a group element 
and Riemann-Hilbert problem. The study of this problem for the trigonometric
Gaudin model based on the $osp(1|2)$ Lie superalgebra is in progress. 
However, we propose a formula for the scalar products of the Bethe states 
which is analogous to the $sl(2)$ case.

The paper is organized as follows. In Section 2 we review main data 
of the quantum trigonometric $osp(1|2)$ spin system: the $osp(1|2)$  
solution to the graded Yang-Baxter equation ($R$-matrix), monodromy 
matrix $T(\l)$, the transfer matrix $t(\l) =  \mathrm {str} \; T(\l)$, its 
eigenvalues and the Bethe equations. The eigenvectors of this
quantum integrable spin system can be constructed only by a complicated recurrence 
procedure \c{T} which is not given here. Nevertheless it is useful to 
remind the main data of the quantum integrable spin system because
some characteristics of the corresponding Gaudin model can be obtained
easily as a quasi-classical limit of these data. The trigonometric 
$osp(1|2)$ Gaudin model and its creation operators $B_M$ are discussed thoroughly 
in Section 3. Some of the most important properties of these operators are 
formulated and demonstrated pure algebraically: antisymmetry with respect 
to their arguments, commutation relations with the trigonometric loop superalgebra 
generators, commutation relations with the generating function $t(\l)$ of the Gaudin 
Hamiltonians, a differential identity, valid in the case of the Gaudin 
realization of the loop superalgebra. Using these properties of the 
$B$-operators we prove in Section 4 that acting on the lowest spin vector 
$\Om _-$ these operators generate eigenvectors of the generating function 
of integrals of motion, provided the Bethe equations are imposed on the 
arguments of the $B$-operators. Possible modifications of the Gaudin Hamiltonians 
are pointed out, also. In particular, one of them yields Ricardson type Hamiltonian. 
An algebraic proof is given in Section 5 that constructed Bethe vectors are entering 
into solutions of the Knizhnik-Zamolodchikov equation of conformal field theory. 
Quasi-classical asymptotic with respect to a parameter of the Knizhnik-Zamolodchikov 
equation permits us to calculate the norm of the eigenstates of the Gaudin Hamiltonian.
We pointed out that modification of Gaudin Hamiltonians by adding Cartan element 
requires more complicated change of KZ system, as opposed to the rational case.
Further development on possible evaluation of correlation functions 
is discussed in Conclusion. Finally, some definitions of the orthosymplectic 
Lie superalgebra $osp(1|2)$ are given in the Appendix.
 

\section{Quantum $osp(1|2)$ superalgebra and corresponding spin system}
 
The quantum superalgebra $\mathcal{U} _q \left (osp(1|2)\right)$ as 
a deformation of the universal enveloping algebra of the Lie superalgebra 
$osp(1|2)$ (see Appendix), is generated by three elements $h, v^+, v^-$ 
\c{KR,HS}. The q-deformed commutation relations between the generators are
\be \la{q-osp}
\left[ h, v^{\pm} \right] _- = \pm v^{\pm} \;, \quad   
\left[ v^+, v^- \right] _+ = - \f{q^{h}-q^{-h}}{q - q^{-1}} := - [h] _q \;.
\ee
Its center is spanned by the q-deformed Casimir element
\be \la{q-Cas}  
\!\!\!c_2 (q) = A (q) \left( [h] _q \right) ^2 + B (q)
- \f {\left( q ^{1/2} + q ^{-1/2} \right) ^2}{2} 
\left [ {v ^+} ^2 , {v ^-} ^2 \right ] _+
+ \f {\left( q + q ^{-1} \right)}{4} \left( q ^{h} + q ^{-h} \right)
\left [ v ^+ , v ^- \right ] _-  
\ee
with $2 A (q) = q + q ^{-1} + \half ( q - q ^{-1} ) ^2$ and 
$B (q) = ( q^{1/2} + q^{-1/2} ) ^{-2}$. The q-deformed Casimir element
can be written also in the form
\be \la{q-Cas-}
c_2 (q) = \left( [h-1/2] _q \right) ^2
- \left( q^{1/2} + q^{-1/2} \right) ^2 {v ^+} ^2 {v ^-} ^2
+ \left( q^{h-1} + q^{-h+1} \right) v^+ v^- \,. 
\ee
Like in the case of the Lie superalgbra $osp(1|2)$ (see Appendix), there exists an
element
\be \la{q-c1} 
c_1 (q) =  \left( [ h - 1/2 ] _q + 
\left( q^{1/2} + q^{-1/2} \right) v^+ v^- \right) \, , \quad 
\left( c_1 (q) \right) ^2 = c _2 (q) \,,
\ee
with a grading property $[c_1 , h] = 0$ and $c _1  v^{\pm} = -  v^{\pm}c _1$.
In the quasi-classical limit $q \to 1$ the relations (2.1-4) are reduced to the
Lie superalgebra  $osp(1|2)$ ones \c{Ritt}. There is a coproduct map 
$\D : \mathcal{U} _q \to \mathcal{U} _q \otimes \mathcal{U} _q$
consistent with the commutation relations \Ref{q-osp} and a universal R-matrix 
$\mathcal {R}$ as an element of $\mathcal{U} _q \otimes \mathcal{U} _q$
\c{KR,HS}. Let us write its matrix form $R (q) = (\rho \otimes \rho) \mathcal {R}$ 
in the basis $e_1\otimes e_1, e_1\otimes e_2, e_1\otimes e_3, \ldots , e_3\otimes e_3$ 
of the tensor product of two copies of the fundamental representation 
$V^{(1)}\otimes V^{(1)}$, $\rho : \mathcal{U} _q \to \mathrm{End} \; (V^{(1)})$,
\baa \la{Rmat} 
R (q) = \left( \ba{ccccccccc} 
q &  & & & & &  & & \\
&1&  &a & & &  & & \\
& &q^{-1}& &b& &c& & \\
& &  &1&   & &  & & \\
& & & & 1 & &b& & \\
& &  & &   &1&  &a& \\
& & & & & &q^{-1}& & \\
& &  & &   & & &1& \\
& &  & &   & &  & &q \\
\ea \right) \,,
\eaa
here $a=q - q^{-1}$, $b=q^{-3/2} - q^{1/2}$ and $c=(1+q^{-1})(q - q^{-1})$. 
Multiplying $R (q)$ by the graded permutation $\mathcal{P}$ of
$V^{(1)}\otimes V^{(1)}: (\mathcal{P})_{ab;cd} = (-1) ^{p(a)p(b)} \d _{ad} \d _{bc}$,
one gets the braid group form of the R-matrix $\check{R} (q) = R(q) \mathcal{P}$ 
which has a spectral decomposition. Using the projectors on the irreducible 
representation components in the Clebsch-Gordan decomposition 
$V ^{(1)}\otimes V ^{(1)} = V ^{(2)}\oplus V ^{(1)}\oplus V ^{(0)}$
one can represent this R-matrix in the form \c{KR, HS}
\begin{equation}
\check{R} (q) = - q ^{-2} P _0 (q) - q ^{-1} P_1 (q) + q P _2 (q)
\end{equation}
where projectors are
\baa \la{P0}
P _0 (q) = \f {q^{1/2}+q^{-1/2}} {q^{3/2} + q^{-3/2}}
\left( \ba{ccccccccc}
0& & & & & & & & \\
&0 & & & & & & & \\
& &q^{-1} & &q^{-1/2}& &-1& & \\
& & &0 & & & & & \\
& &-q^{-1/2}& &-1& &q^{1/2}& & \\
& & & & &0 & & & \\
& &-1& &-q^{1/2}& &q& & \\
& & & & & & &0& \\
& & & & & & & &0
\ea \right) \,,
\eaa

\baa \la{P1}
\!\!\!P _1  (q) = \f {1} {q + q^{-1}}
\left( \ba{ccccccccc}
0& & & & & & & & \\
&q^{-1}& &-1 & & & & & \\
& &-1& &- {\nu} & &1& & \\
&-1&  &q&   & &  & & \\
& & {\nu} & & {\nu} ^2& &- {\nu} & & \\
& &  & &   & q^{-1}& &-1 & \\
& &1& & {\nu} & &-1 & & \\
& &  & &   &-1&  &q & \\
& &  & &   & &  & & 0
\ea \right) \,,
\eaa
here $\nu = q^{1/2}+q^{-1/2}$, and $P_2 (q) = I -  P _0 (q) - P_1 (q)$. 
By the Baxterization procedure, or simply changing functions 
$(\l - a)/(\l + a)$ to the trigonometric ones 
$\sinh (\l - a) / \sinh (\l + a)$ in the $osp(1|2)$- invariant R-matrix
\c{K85, KM} one gets the trigonometric $R$-matrix related to the quantum 
affine algebra $\mathcal {U} _q (\widehat{osp(1|2)})$
\be \la{Rtrig}
\check R ( \l , \eta ) =
P _2 - \f {\sinh ( \l - 2 \eta )}{\sinh ( \l + 2 \eta )} P _1 -
\f {\sinh ( \l - 3 \eta )}{\sinh ( \l + 3 \eta )} P_0 \, ,
\ee
here $q = e ^{2 \eta}$ (normalized to $\l = 0$ $\check {R} = \mathbf {1})$.
The $L$-operator of the quantum spin system on a one-dimensional lattice 
with $N$ sites coincides with $R$-matrix acting on a tensor product 
$V_0 \otimes V_a$ of auxiliary space $V_0$  and 
the space of states at site $a = 1, 2,\dots N$ 
\be \la{Lop-q}  
L_{0a}(\l - z_a) = R_{0a} (\l - z_a) \,, 
\ee
where \(z_a\) is a parameter of inhomogeneity (site dependence) and 
$R_{0a} (\l , \eta ) = \mathcal {P}  \check {R} ( \l , \eta )$ is the usual R-matrix. 
Corresponding monodromy matrix $T$ is an ordered product of the $L$-operators 
\be \la{T-q}
T(\l ; \{z_a\}_1^N) = L_{0N}(\l - z_N) \dots  
L_{01}(\l - z_1) = \underset{\longleftarrow}{\prod_{a=1}^{N}}
L_{0a} (\l - z_a) \,.  
\ee
The commutation relations of the $T$-matrix entries follow form the 
($Z_2$-graded) FRT-relation \c{KS}
\be \la{RTT}
R_{12}(\l - \mu)T_1(\l)T_2(\mu) = 
T_2(\mu)T_1(\l)R_{12}(\l - \mu) \,.  
\ee 
Multiplying \Ref{RTT} by $R_{12}^{-1}$ and taking the super-trace 
over $V_1 \otimes V_2$, one gets commutativity of the transfer 
matrix 
\be \la{tq} 
t(\l) = \mathrm{str} \; T (\l) = \sum_j (-1)^{j+1}T_{jj}(\l ; \{z_a\}_1^N)
= T_{11} - T_{22} + T_{33}  
\ee 
for different values of the spectral parameter 
$t(\l)t(\mu) = t(\mu)t(\l)$.

The choice of the $L$-operators \Ref{Lop-q} corresponds to 
the following space of states of the $osp(1|2)$-spin system  
\ben
\mathcal {H} = \underset {a=1}{\overset{N}{\otimes}} V_a^{(1)} \;. 
\een
The eigenvalues of the transfer matrix $t(\l)$ in this space 
are \c{K85,M,LS} 
\baa 
\la{lq} 
\L (\l ; \{\mu_j\}_1^M) &=& 
\a_1^{(N)}(\l;\{z_a\}_1^N) \prod_{j=1}^M S_1(\l - \mu_j) 
-\a_2^{(N)}(\l; \{z_a\}_1^N) \times
\non
&\times& \prod_{j=1}^M S_1 \left( \l - \mu_j + {\eta} \right) 
S_{-1}(\l - \mu_j + 2 \eta ) + 
\non
&+&  \a _3^{(N)}(\l; \{z_a\}_1^N) \prod_{j=1}^M 
S_{-1} \left( \l - \mu_j + 3 {\eta} \right) \,, 
\eaa
where $\a_j^{(N)}(\l;\{z_a\}_1^N)=\prod_{b=1}^N 
\a_j(\l - z_b)\,; j = 1, 2, 3\,,$  
\baa
\a_1(\l) &=& \sinh (\l+ 2 \eta) \sinh (\l+ 3 \eta) \,,   
\quad \a_2(\l) = \sinh (\l) \sinh (\l + 3 \eta) \,,  
\non
\a_3(\l) &=& \sinh (\l) \sinh (\l + \eta) \,,  
\quad S_n(\mu) = \f {\sinh (\mu - n \eta)}{\sinh (\mu + n \eta)} \;. 
\eaa 
Although according to \Ref{lq} the eigenvalue has formally two 
sets of poles at $\l = \mu_j - \eta$ and $\l = \mu_j - 2 \eta$, the 
corresponding residues are zero due to the Bethe equations on the 
parameters $\{ \mu _j \}$ of the eigenstate \c{K85,M,LS} 
\be \la{BEq} 
\prod_{a=1}^N
\f{\sinh (\mu_j -z_a +  \eta)}{\sinh (\mu_j -z_a - \eta)} = 
\prod_{k=1}^M S_1(\mu_j - \mu_k) S_{-2}(\mu_j - \mu_k) \;.
\ee

If we take different spins $l_a$ at different sites of the lattice 
and the following space of states 
\ben
{\cal H} = \underset {a=1}{\overset{N}{\otimes}} V^{(l_a)}_a \;, 
\een
then the factors on the left hand side of \Ref{BEq} will 
be spin dependent too.

Due to the more complicated structure of the  $R$-matrix \Ref{Rtrig} 
(see \Ref{P0}, \Ref{P1}) than the $gl(n)$, or $gl(m|n)$ trigonometric $R$-matrices, 
the commutation relations of the entries $T_{ij}(\l)$  of the $T$-matrix \Ref{T-q} 
have more terms and construction of the eigenstates of the transfer matrix 
$t(\l)$ by the algebraic Bethe Ansatz can be done only using a complicated 
recurrence relation expressed in terms of $T_{ij}(\mu_k)$ \c{T}. It will be 
shown below that due to a simplification of this recurrence relation in the 
quasi-classical limit $\eta \to 0$ one can solve it and find the creation operators 
for the trigonometric $osp(1|2)$ Gaudin model explicitly. Furthermore, the 
commutation relations between the creation operators and the generators of the 
trigonometric loop superalgebra as well as the generating function $t(\l)$ 
of the Gaudin Hamiltonians will be given explicitly, yielding the solution
to the eigenvalue problem.


\section{$osp(1|2)$ trigonometric Gaudin model}

As in the case of any simple Lie algebra, the trigonometric classical 
$r$-matrix of the orthosymplectic Lie superalgebra $osp(1|2)$ 
can be expressed in a pure algebraic form as an element in the tensor product 
$osp(1|2) \otimes osp(1|2)$
\baa \la{crm}
{\hat r} \left( \l \right) &=& \coth (\l) \, h \otimes h + \f {2}{\sinh (\l)} 
\left( e ^{- \l} X^+ \otimes X^- 
+ e ^{\l} X^- \otimes X^+ \right) \non
&+& \f {1}{\sinh (\l)} \left( e ^{- \l} 
v^+ \otimes v^- - e ^{\l} v^- \otimes v^+ \right) \;,
\eaa
and it is a solution of the $Z_2$-graded classical YBE \Ref{cYB} \c{KS}.
This $r$-matrix can be decomposed naturally into positive and negative
parts \c{STS}
\be \la{crm-triang}
{\hat r} ( \l ) = \f 1{\sinh \l} \left( e ^{\l} r ^{(-)} +
e ^{-\l} r ^{(+)} \right)
\ee
\ben
\!\!\!=\f {e ^{\l}}{\sinh \l} \left ( \f 12 h \otimes h 
+ 2 X^- \otimes X^+ -  v^- \otimes v^+ \right)
+ \f {e ^{-\l}}{\sinh \l} \left ( \f 12 h \otimes h + 2 X^+ \otimes X^-
+ v^+ \otimes v^- \right) .
\een
It can also be represented in another form useful for modifications
\be \la{crm-casimir}
{\hat r} ( \l ) = \coth ( \l ) \; c ^{\otimes}_2 + 
2 \left(X^- \otimes X^+ - X^+ \otimes X^- \right) 
- \left(v^- \otimes v^+ + v^+ \otimes v^- \right) \;, \\
\ee
here
\be
\la{cross-casimir}
c_2^{\otimes} = h \otimes h + 2 \left( X^+ \otimes X^- + X^- \otimes X^+ \right) 
+ \left(v^+ \otimes v^- - v^- \otimes v^+ \right) \;.
\ee
The matrix form of $\hat r$ in the fundamental representation 
of $osp(1|2)$ follows from \Ref{crm} by substituting appropriate 
$3 \times 3$ matrices instead of the $osp(1|2)$ generators and taking 
into account $Z_2$-graded tensor product of even and 
odd matrices. One can get it also as the quasi-classical limit
$\eta \to 0$ from the $R$-matrix \Ref{Rtrig}.
Let us write explicitly the matrix form of $\hat r$ 
in the basis of the tensor product of two copies of the fundamental 
representation $V ^{(1)}\otimes V ^{(1)}$ (see Appendix)
\baa \la{r}  
\!\!\!r (\l) = \f {1}{\sinh(\l)} \left( \ba{ccccccccc}
\cosh(\l) &  & & & & &  & & \\
&0&  & e^{-\l} & & &  & & \\
& &-\cosh(\l)& &- e^{-\l} & & 2e^{-\l}& & \\
&e^{\l}&  &0&   & &  & & \\
& & e^{\l}& & 0 & &-e^{-\l}& & \\
& &  & &   &0&  &e^{-\l}& \\
& & 2e^{\l}& & e^{\l} & &-\cosh(\l)& & \\
& &  & &   &e^{\l}&  &0& \\
& &  & &   & &  & &\cosh(\l) 
\ea \right) \non
\eaa
with all the other entries of this $9\times 9$ matrix being
identically equal to zero. 

A quasi-classical limit  $\eta \to 0$ of the FRT-relations \Ref{RTT} 
\( \left( R(\l; \eta) = I + \eta r(\l) + {\cal O}(\eta^2)  \right. \) and
\( \left. T(\l; \eta) = I + \eta L(\l) + {\cal O}(\eta^2) \right) \)
results in a matrix form of the loop superalgebra relation, 
the so-called  Sklyanin linear bracket,  
\be
\left[ \underset 1 {L}(\l), \, \underset 2 {L} (\mu) \right] = - 
\left[ r_{12}(\l - \mu) \, , \, \underset 1 {L}(\l) + \underset 2 {L}(\mu) \right] \,. \la{rL} 
\ee
Both sides of this relation have the usual commutators of even
$9 \times 9$ matrices $\underset 1 {L}(\l)  = L(\l) \otimes I_3$,
$\underset 2 {L}(\mu)  = I_3 \otimes L(\mu)$ and $r_{12}(\l - \mu)$, where $I_3$ is 
$3 \times 3$ unit matrix and $L(\l)$ has loop superalgebra valued entries: 
\baa \la{L}  
L(\l) = \left( \ba{ccc}
h(\l) & - v ^-(\l) & 2 X ^{-}(\l) \\
v ^+(\l) & 0 & v ^-(\l) \\
2 X ^+(\l) & v ^+(\l) & - h(\l) 
\ea \right) \,.
\eaa
From the expression \Ref{crm-triang} of the classical $r$-matrix it is natural to
assume that the $L$-operator has a triangular decomposition $L _{\pm}$ as 
$\l \to \pm \infty$
\baa \la{L+}  
L_+ &=& h \otimes h_{gl} + 4 X^- \otimes X^+_{gl} - 2 v^- \otimes v^+_{gl} \;,\\
\la{L-}  
L_- &=& h \otimes h_{gl} + 4 X^+ \otimes X^-_{gl} + 2  v^+ \otimes v^-_{gl} \;.
\eaa
Here the first factors are generators in the fundamental representation $V ^{(1)}$
(see Appendix) and the second factors are generators of a finite dimensional
$osp(1|2)$ Lie superagebra. 

The relation \Ref{rL} is a compact matrix form of the following commutation 
relations between the generators $h(\l), \, v^{\pm}(\mu), \, X^{\pm}(\nu)$ 
of the trigonometric loop superalgebra $\mathcal {L}_t (osp(1|2))$
\baa \la{Ga}
\left[h(\l) \; , \; h(\mu) \right]_- &=& 0 \,, \non 
\left[h(\l) \; , \; X^{\pm}(\mu) \right]_- &=& \f {\pm \,2}{\sinh(\l -\mu)} 
\left( \cosh(\l -\mu) X^{\pm}(\mu) - e^{\mp (\l-\mu)} X^{\pm}(\l) 
\right) \,, \non 
\left[X ^+(\l) \; , \; X ^-(\mu) \right]_- &=& \f {- \, e^{(\l-\mu)}}{\sinh(\l -\mu)}
\left( h(\l) - h(\mu) \right)		    \,,  \non 
\left[h(\l) \; , \; v^{\pm}(\mu) \right]_- &=& \f {\pm \,1}{\sinh(\l -\mu)}
\left( \cosh(\l -\mu) v^{\pm}(\mu) - e^{\mp (\l-\mu)} v^{\pm}(\l) 
\right) \,, \non  
\left[v^+(\l)\; , \; v^-(\mu) \right]_+ &=& \f {e^{(\l-\mu)}}{\sinh(\l -\mu)}
\left( h(\l) - h(\mu) \right)		    \,,  \non
\left[v^{\pm}(\l)\; , \; v^{\pm}(\mu) \right]_+ &=& \f {\pm \,2}{\sinh(\l -\mu)}
\left( e^{\pm (\l -\mu)} X^{\pm}(\mu) -  e^{\mp (\l -\mu)} X^{\pm}(\l) \right) 
\,,  \non 
\left[X^{\pm}(\l) \; , \; v^{\mp}(\mu) \right]_- &=& \f {e^{\pm \, (\l-\mu)}}{\sinh(\l -\mu)}
\left( v^{\pm}(\mu) - v^{\pm}(\l) \right) \,,  \non 
\left[X^{\pm}(\l) \; , \; v^{\pm}(\mu) \right]_-  &=& \left[X^{\pm}(\l) \; , \; X^{\pm}(\mu) \right]_- 
= 0  \,.  \non
\eaa

In order to define a dynamical system besides the algebra of observables
we need to specify a Hamiltonian. Due to the $r$-matrix relation 
\Ref{rL} the elements 
\baa \la{tG} 
t(\l) &=& \frac 12 \; \mathrm{str} \; L^2(\l) = h^2(\l) + 
2 [X^+(\l)\,,\,X^-(\l)]_+ + [v^+(\l)\,,\,v^-(\l)]_- 
\non 
&=&h^2(\l) + h'(\l) + 4 X^+(\l)X^-(\l) + 2 v^+(\l)v^-(\l) 
\eaa 
commute for different values of the spectral parameter 
\baa \la{tcom} 
t(\l) t(\mu) = t(\mu) t(\l) \,. 
\eaa 
Thus, $t(\l)$ can be considered as a generating function of integrals of motion. 
The supertrace in \Ref{tG} for an even matrix $\{A_{ij}\}$ means
$\mathrm{str} A = \sum _{i=1}^3 (-1) ^{p(i)}  A _{ii}$, and we use the grading
$p(1) = p(3) = 0$, $p(2)=1$ (see Appendix).

One way to show \Ref{tcom} is to notice that the commutation relation between 
$t(\lambda)$ and $L(\mu)$ can be written in 
the form
\baa \la{LM} 
\left[ t(\l) \,,\, L (\mu) \right] = 
\left[ M(\l , \mu ) \, , \, L (\mu) \right] \,, 
\eaa 
where 
\baa \la{M-op} 
M(\l , \mu ) = - \underset 1 {\str} \left(  
r_{12}(\l - \mu) \underset 1 {L}(\l) \right) - \frac 12
\underset 1 {\str} \left(  r_{12}^2 (\l - \mu)  \right) \,,
\eaa 
and the second term is a quantum correction, which is absent if we consider 
the left hand side of \Ref{rL}, \Ref{Ga} as Poisson brackets \c{FTakh}.
Also, in the $sl(n)$ case this term does not contribute since
is proportional to the unit matrix. Using \Ref{r} and \Ref{L} 
it is straightforward to calculate $M(\l , \mu )$ explicitly
\baa \la{Mexpl}  
M(\l , \mu ) &=& \f {-2}{\sinh (\l - \mu )} 
\left( \ba{ccc}
\cosh (\l - \mu ) \, h(\mu ) & - e ^{\l - \mu} \, v ^-(\mu) & 2 e ^{\l - \mu} \, X ^{-}(\mu) \\
e ^{-(\l - \mu)} \, v ^+(\mu) & 0 & e ^{\l - \mu} \, v ^-(\mu) \\
2 e ^{-(\l - \mu)} \, X ^+(\mu) & e ^{-(\l - \mu)} \, v ^+(\mu) & - \cosh (\l - \mu ) \, h(\mu) \ea \right)  
\non
\non
&-& \f {1}{\sinh ^2 (\l - \mu )} 
\left( \ba{ccc}
\cosh ^2 (\l - \mu ) + 1 & 0 & 0 \\
0 & -2 & 0 \\
0 &  0 & \cosh ^2 (\l - \mu ) + 1 \ea \right) \,.
\eaa
Substituting \Ref{Mexpl} into the equation \Ref{LM} we obtain the commutation relations
between $t(\l)$ and the generators of the superalgebra $\mathcal {L}_t (osp(1|2))$. In particular
\baa
\la{tX}
\left[t(\l) \; , \; X^+(\mu) \right]_- &=& 4 \coth (\l - \mu) \, X^+(\mu) h (\l)
- \f {4 e ^{-(\l - \mu)}}{\sinh (\l - \mu)} \, X^+(\l) h (\mu) + 4  X^+(\mu) \non
&-& \f {2 e ^{-(\l - \mu)}}{\sinh (\l - \mu)} \left( v^+ (\l) v^+ (\mu) 
- v^+ (\mu)  v^+ (\l) \right) \,,
\eaa
\baa
\la{tv}
\!\!\!\left[t(\l) \; , \; v^+(\mu) \right]_- &=& 2 \coth (\l - \mu) \, v ^+ (\mu) h (\l)
- \f {2 e ^{-(\l - \mu)}}{\sinh (\l - \mu)} \, v ^+ (\l) h (\mu) + v^+(\mu) \non
&+& \f {4}{\sinh (\l - \mu)} \left( e ^{-(\l - \mu)} \,  X^+(\l)v^-(\mu) - 
e ^{-(\mu - \l)} \, X^+(\mu)v^-(\l) \right) \,. \non
\eaa

Preserving some generality we can consider the representation space
\({\cal H} \) of the dynamical algebra to be a lowest spin 
\(\rho (\l)\) representation of the loop superalgebra with the 
lowest spin vector \(\Omega_-\) 
\begin{equation} 
\la{vac} 
h (\l) \Om_- = \rho (\l) \Om_- \,, \quad v^-(\l) \Om_- = 0 \,. 
\end{equation} 
One can study spectrum and eigenstates of $t(\l)$ in this general representation 
\({\cal H} \). However, to have a physical interpretation we will use a 
local realization of the trigonometric superalgebra $\mathcal {L} _t (osp(1|2))$
with
\ben
\mathcal {H} = \underset {a=1}{\overset{N}{\otimes}} V _a \;, 
\een
as a tensor product of $osp(1|2)$ representations. Then,
\begin{equation} \la{Gr-h}
h(\l)=\sum_{a=1}^N \coth(\l-z_a)  \, {h_a} 
\end{equation}
\baa \la{Gr-v}
v ^+(\l) = \sum_{a=1}^N \f {e^{\l-z_a}}{\sinh (\l-z_a)} \, v^+_a 
\,, && \quad 
v ^-(\l) = \sum_{a=1}^N \f {e^{-\l+z_a}}{\sinh (\l-z_a)} \, v^-_a \,, \\
\la{Gr-X}
X ^+(\l) = \sum_{a=1}^N \f {e^{\l-z_a}}{\sinh (\l-z_a)}  \, X ^+_a 
\,, && \quad  
X ^{-}(\l) = \sum_{a=1}^N \f {e^{-\l+z_a}}{\sinh (\l-z_a)}  \, X ^-_a \,,
\eaa
where $h _a, v^{\pm}_a, X^{\pm}_a \in \mathrm{End} \, (V_a)$ 
are \(osp (1|2) \) generators in a representation $V_a$ associated with each 
site $a$. If in this realization one considers the limits $\l \to \pm \infty$ 
then one finds the expressions of the generators of the global Lie superalgebra 
$osp(1|2)\subset \mathcal{L}_t (osp(1|2))$ in terms of the local generators
\be \la{gl-gen}
Y_{gl} = \sum _{a=1}^N Y _a \;,
\ee
here $Y = ( h , X ^{\pm} , v ^{\pm})$.

In particular, a representation of the Gaudin realization can be 
obtained by considering at each site $a$ an irreducible representations 
$V_a^{(l_a)}$ of the Lie superalgebra \(osp(1|2)\) defined by a spin $l_a$ 
and a lowest spin vector $\o_a$ such that $v_a^-\o_a = 0$ 
and $h_a \o_a = -l_a \o_a$. Thus,
\begin{equation} \la{Gre}
\Om _- = \mathop{\otimes} \limits_{a=1}\limits^N \o _a \;, \quad 
\hbox {and} \quad 
\rho (\l) = \sum_{a=1}^N  (-l_a) \, \coth({\l - z_a}) \;.
\end{equation}

It is a well-known fact in the theory of Gaudin models \c{GB,S87} that the Gaudin 
Hamiltonian is related to the classical $r$-matrix \Ref{Gh}, \Ref{crm-triang}
\baa \la{sGh}
H^{(a)} &=& \sum_{b \neq a} r_{ab}(z_a - z_b) = \sum_{b \neq a} \f 1 {\sinh (z_a-z_b)} 
\left( e ^{(z_a-z_b)} r ^{(-)}_{ab} + e ^{(z_b-z_a)} r ^{(+)}_{ab} \right) \non
&=& \sum_{b \neq a} \coth (z_a - z_b) {h_a} {h_b} + \f {2}{\sinh (z_a - z_b)}
\left( e ^{-(z_a - z_b)} X^+_a X^-_b + e ^{(z_a - z_b)} X^-_a X^+_b \right) \non
&+& \f {1}{\sinh (z_a - z_b)} \left( e ^{-(z_a - z_b)} 
v^+_a v^-_b - e ^{(z_a - z_b)} v^-_a v^+_b \right)
\,, 
\eaa
and can be obtained as the residue of the operator $t(\l)$ at the point $\l = z_a$ 
using the expansion 
\baa \la{tpole}
t(\l) = h ^2_{gl} + \sum_{a =1}^N \left( \f {c_2 (a)}{\sinh ^2 (\l - z_a)}
+ \, 2 \, \f {e ^{\l -z_a}}{\sinh (\l - z_a)}  \, H^{(a)} \right) \,, \quad
\eaa
$c_2 (a) = {h _a} ^2 + 2 \left( X ^+_a X ^-_a 
+ X ^-_a X ^+_a \right) + v  ^+_a v ^-_a - v ^-_a v ^+_a$. As opposed to the rational 
$osp (1|2)$-invariant case \c{KM}, the generating function \Ref{tpole} commutes
only with one generator $h _{gl}$ of the global superalgebra $osp(1|2)$
\be
\left[ t(\l) , h _{gl} \right] = 0 \;.
\ee

To construct the set of eigenstates of the generating function 
of integrals of motion $t(\l)$ we have to define appropriate 
creation operators. The creation operators used in the $sl(2)$ 
Gaudin model coincide with one of the $L$-matrix entry \c{GB,S87}. 
However, in the \(osp(1|2)\) case the creation operators are 
complicated functions of the two generators of the loop superalgebra 
$X^+(\l)$ and $v^+(\mu)$.

\begin{definition}
Let $B_{M}(\mu_1, \dots , \mu_M)$ belong to the Borel subalgebra
of the \(osp (1|2) \) loop superalgebra ${\mathcal L}_t(osp (1|2))$
such that
\baa \la{recr}  
B_{M}(\mu_1, \dots , \mu_M) = v^+(\mu_1) B_{M-1}(\mu_2, \dots , \mu_M)  \non
+ 2 X^+(\mu_1) \sum_{j=2}^{M} (-1)^j \f {e ^{-(\mu_1 - \mu_j)}}
{\sinh (\mu_1 - \mu_j)} B_{M-2}^{(j)}(\mu_2, \dots , \mu_M)
\eaa
with $B_0 = 1,$ $B_1(\mu) = v^+(\mu) $ and $B_M = 0$ for $M < 0$. 
The notation
$B_{M-2}^{(j)}(\mu_2, \dots ,\mu_M)$ means that the argument 
$\mu_j$ is omitted.
\end{definition}
As we will show below, the $B$-operators are such that the Bethe 
vectors are generated by their action on the lowest spin vector 
$\Om _-$ \Ref{vac}. To prove this result we will need some 
important properties of the $B$-operators. All the properties of 
the creation operators $B_M(\mu_1, \dots , \mu_M)$ listed below can 
be demonstrated by induction method. Since the proofs are lengthy 
and quite technical we will present only two of them.

\begin{lemma}
The creation operators $B_{M}(\mu_1, \dots , \mu_M)$ are 
antisymmetric functions of their arguments
\baa \la{asym} 
B_{M}(\mu_1, \dots , \mu_{k},  \mu_{k+1}, \dots , \mu_M) &=& 
- \; B_{M} (\mu_1, \dots , \mu_{k+1}, \mu_{k}, \dots , \mu_M) 
\,, \non
\eaa
here $1\leq k < M$ and $M \ge 2$. 
\end{lemma}

\begin{lemma}
The commutation relations between the creation operator $B_M$
and the generators $v^+(\l)$,  $h(\l)$, $v ^-(\l)$ of the loop 
superalgebra are given by 
\baa \la{vB} 
v^+(\l) B_{M} &=& (-1)^M B_{M} v^+(\l) + 2 \sum_{j=1}^M 
\f {(-1)^j} {\sinh (\l - \mu _j)} \non
&&\times \left( e^{-(\l - \mu _j)} X^+(\l) - e^{-(\mu _j - \l)} 
X^+(\mu _j) \right) B_{M-1}^{(j)} \,,
\eaa
\baa \la{hB} 
h(\l) B_{M} &=& B_{M} \left( h(\l) + \sum_{i=1}^M {\coth (\l - \mu_i)} \right)
+ \sum _{i=1}^M (-1)^i \f{e^{-(\l - \mu _i)}}{\sinh (\l - \mu _i)} \times \non
&&\times \left( v ^+(\l) B_{M-1}^{(i)} + 2 X ^+ (\l) \sum_{j \neq i}^M (-1)^{j+\Th (i-j)}
\f{e^{-(\mu _i - \mu _j)}}{\sinh (\mu _i -\mu _j)} B_{M-2}^{(i,j)} \right) \quad
\eaa
\baa \la{v-B}
v ^-(\l)  B_{M} &=& (-1) ^M B_{M} v ^-(\l) + \sum _{j=1}^M 
(-1)^{j-1} B_{M-1}^{(j)}
\left( \f{e ^{-(\l-\mu _j)}} {\sinh (\l - \mu _j)} 
\left( h (\l) - h (\mu _j) \right)  \right. \non
&&\left. + 
\sum_{k \neq j}^M \f{e ^{-(\l-\mu _k)}}{\sinh (\l - \mu _k)}
\f{e ^{-(\mu _k-\mu _j)}}{\sinh (\mu _k - \mu _j)} \right ) 
+ v ^+(\l) \sum_{i < j}^M (-1) ^{i-j-1}  \non
&&\times \f {e^{-(\mu _i - \mu _j)}}{\sinh (\mu _i - \mu _j)}B^{(i,j)}_{M-2}
\left( \f {e ^{-(\l-\mu _i)}} {\sinh (\l - \mu _i)} 
+ \f {e ^{-(\l-\mu _j)}} {\sinh (\l - \mu _j)} \right) \,.
\eaa
here the upper index of $B_{M-1}^{(j)}$ means that the argument 
$\mu_j$ is omitted, the upper index of $B_{M-2}^{(i,j)}$ means 
that the parameters $\mu_i, \mu_j$ are omitted and $\Th (j)$ 
is Heaviside function \[ 
\Th (j) = \left\{ \ba{ll}
                1 & \mbox {if $j > 0$} \\
		0 & \mbox {if $j \leq 0$}
		 \ea \right. \;.
\]
\end{lemma}

It is useful to have explicit formulas for the commutators between the 
global generators and the $B$-operators.

\begin{remark}
The commutation relations between the generators of the global \(osp (1|2) \) 
\Ref{gl-gen} and the $B_M$ elements follow from the previous lemma 3.2. To see this 
we take the appropriate limit $\l \to \pm \infty$ in \Ref{vB}, \Ref{hB}, \Ref{v-B}. 
In this way we obtain 
\baa \la{vglB} 
\!\!\!v^+_{gl} B_{M} &=& (-1)^M B_{M} v^+_{gl} - 2 \sum_{j=1}^M (-1)^j
X^+(\mu _j) B_{M-1}^{(j)} \;, \\
\la{hglB}
\!\!\!h _{gl}  B_{M} &=& B_{M} \left( h _{gl}  + M \right) \;, \\
\la{v-glB} 
\!\!\!v^- _{gl} B_{M} &=& (-1)^M  B_{M} \, v^-_{gl} + \sum_{j=1}^M (-1)^j B_{M-1}^{(j)} 
\left( h _{gl} + h(\mu_j) + \sum_{k \neq j}^M 
\f {e ^{(\mu _j - \mu _k)}}{\sinh (\mu _j - \mu _k)} \right) \quad
\eaa 
\end{remark}

\begin{lemma}
The generating function of integrals of motion \(t (\l)\) 
\Ref{tG} has the following commutation relation with 
the creation operator $B_M (\mu_1, \dots , \mu_M)$
\ben \la{tB}
\!\!\!\!\!\!t(\l)B_{M} = B_{M} t(\l) +  B_{M} \left( 2 h (\l) 
\sum _{i=1}^M \coth {(\l -\mu _i)} + 2 \sum_{i < j}^M 
\coth {(\l -\mu _i)} \coth {(\l -\mu _j)} + M \right)
\een
\ben
+ 2 \sum _{i=1}^M (-1)^i \f {e^{-(\l - \mu _i)}}{\sinh (\l - \mu _i)} \left( v ^+(\l) 
B_{M-1}^{(i)} + 2 X ^+ (\l) \sum_{j\neq i}^M (-1)^{j+\Th (i-j)}
\f {e^{-(\mu _i - \mu _j)}}{\sinh (\mu _i -\mu _j)} 
B_{M-2}^{(i,j)} \right)
\een
\be
\times \hb _M (\mu _i) + 4 \sum _{i=1}^M (-1)^{i+M} \f {B_{M-1}^{(i)}}{\sinh (\l - \mu _i)}
\left( e^{-(\l - \mu _i)} X ^+ ( \l )  v ^- ( \mu _i ) 
- e^{-(\mu _i - \l)} X ^+ ( \mu _i )  v ^- ( \l ) \right) \;.
\ee
The notation used here for the operator $\hb _M (\mu _i)$ is
$\hb _M (\mu _i) = h(\mu_i) + \sum_{j\neq i}^M  
\coth (\mu_i - \mu_j)$.
\end{lemma}

In the trigonometric Gaudin realization $(3.17-19)$ the creation operators 
$B_M (\mu_1, \dots , \mu_M)$ have some specific analytical 
properties.

\begin{lemma}
The $B$-operators in the Gaudin realization $(3.17-19)$ satisfy 
an important differential identity
\baa \la{derB}
\f {\p}{\p z_a} B _M &=& \sum _{j=1}^M  \f {\p}{\p \mu _j} 
\left( (-1) ^j \f {e^{\mu_j - z_a}}{\sinh (\mu_j - z_a)} \right. \non
&\times& \left.
\left( v ^+_a B^{(j)}_{M-1} 
+ 2 \, X ^+_a \, \sum _{k\neq j}^M (-1) ^{k+\Th (j-k)}
\f {e ^{-(\mu_j - \mu _k)}} {\sinh (\mu_j - \mu _k)}  
B^{(j,k)}_{M-2} \right) \right) \;.
\eaa
\end{lemma}

This identity will be a fundamental step in establishing a 
connection between the Bethe vectors and solutions to the KZ equation.


The proofs of the lemmas are based on the induction method. 
As illustrations, we prove explicitly lemma 3.1 and the formula \Ref{vB}
in lemma 3.2. 

{\it Proof of lemma 3.1.} Consider $M=2$
\ben 
B_2(\mu_1, \mu_2) = v^+(\mu_1) v^+(\mu_2) + \f { 2 e ^{-(\mu_1 - \mu_2)}} 
{\sinh (\mu_1 - \mu_2)}  X^+(\mu_1)\;.
\een
Using the commutation relations \Ref{Ga} it is straightforward to
check that $B_2(\mu_1, \mu_2)$ is antisymmetric
\ben 
B_2(\mu_1, \mu_2) = - B_2(\mu_2 , \mu_1) \;.
\een
Assume $B_{N}(\mu_1, \dots , \mu_N)$ is antisymmetric $N \geq 2$ and
for $N < M$. We have to prove that\break\hfil 
$B_{M}(\mu_1, \dots , \mu_M)$ is antisymmetric also. 

Consider $j \geq 2$, the antisymmetry of $B_{M}(\mu_1, \dots , \mu_M)$
with respect to $\mu _j$ and $\mu _{j+1}$\break\hfill  follows directly form the
recurrence relation \Ref{recr} and our assumption. Namely, the terms 
$B_{M-2}^{(j)}(\mu_2, ...,\mu_M) e ^{-(\mu_1 - \mu_j)}/\sinh (\mu_1 - \mu_j)$ 
and $B_{M-2}^{(j+1)}(\mu_2, ...,\mu_M) e ^{-(\mu_1 - \mu_{(j+1)})}/
\sinh (\mu_1 - \mu_{(j+1)})$ enter with the opposite sign. 

Therefore we only have to show the antisymmetry of $B_{M}(\mu_1, \dots , \mu_M)$
with respect to the interchange of $\mu _1$ and $\mu _2$. To see this
we have to iterate the recurrence relation \Ref{recr} twice and combine
the appropriate terms
\baa \la{asym12}
B_{M}(\mu_1, \dots , \mu_M) &=& \left( v^+(\mu_1) v^+(\mu_2) + \f { 2 e ^{-(\mu_1 - \mu_2)}} 
{\sinh (\mu_1 - \mu_2)}  X^+(\mu_1) \right) B_{M-2}(\mu_3, \dots , \mu_M) \non
&+& 2  v^+(\mu_1) X^+(\mu_2) \sum_{j=3}^{M} (-1)^{j+1}
\f {e ^{-(\mu_2 - \mu_j)}}{\sinh (\mu_2 - \mu_j)} \;
B_{M-3}^{(j)}(\mu_3, \dots , \mu_M) \non
&+& 2  v^+(\mu_2) X^+(\mu_1) \sum_{j=3}^{M} (-1)^j
\f {e ^{-(\mu_1 - \mu_j)}}{\sinh (\mu_1 - \mu_j)} \;
B_{M-3}^{(j)}(\mu_3, \dots , \mu_M) \non
&+& 4  X^+(\mu_1) X^+(\mu_2) \sum_{j=3}^{M} (-1)^j 
\f {e ^{-(\mu_1 - \mu_j)}}{\sinh (\mu_1 - \mu_j)}
\sum_{k=3}^{M} (-1)^{k+{\Th} (j-k)} \times \non
&\times& \f {e ^{-(\mu_2 - \mu_k)}}{\sinh (\mu_2 - \mu_k)}
B_{M-4}^{(j , k)}(\mu_3, \dots , \mu_M) \non
\eaa
where $B_{M-4}^{(j , k)}(\mu_3, \dots , \mu_M)$  means that the arguments 
$\mu_j$ and $\mu _k$ are omitted.
Since $v^+(\mu)$ commutes with $X^+(\nu)$, the antisymmetry of 
the right hand side of \Ref{asym12} with respect to $\mu _1$ and 
$\mu _2$ follows. Hence we have demonstrated the lemma.
{\qed}

{\it Proof of lemma 3.2.} Here we prove explicitly only formula \Ref{vB}.
In particular, when $M=1$ the expression \Ref{vB} 
is just the anticommutator between $v^+(\l)$ and $v^+(\mu)$. 
Using the recurrence relations \Ref{recr} it is straightforward to 
check that the formula \Ref{vB} holds for $M=2$
\baa \la{vB2} 
v^+(\l) B _2 (\mu_1 , \mu_2) &=& B_2 (\mu_1 , \mu_2) v^+(\l) \non
&-& \f 2{\sinh (\l - \mu _1)} \left( e^{-(\l - \mu _1)} X^+(\l) 
- e^{\l - \mu _1} X^+(\mu _1) \right) \; v^+(\mu _2) \non
&+& \f 2{\sinh (\l - \mu _2)} \left( e^{-(\l - \mu _2)} X^+(\l) 
- e^{\l - \mu _2} X^+(\mu _2) \right) \; v^+(\mu _1) \;. \quad
\eaa
Therefore we can proceed to demonstrate the lemma 3.3 by induction.
Assume that the relation \Ref{vB} holds for $B_N$, $M \geq N \geq 2$.
Then we have to show the formula \Ref{vB} is valid for $M+1$. 
We use the recurrence relations \Ref{recr} to write
\baa \la{vBN+1a} 
v^+(\l) B_{M+1} &=& v^+(\l) \left( v^+(\mu_1) B_M + 2 X^+(\mu_1) 
\sum_{j=2}^{M+1} \f {e ^{-(\mu_1 - \mu_j)}}
{\sinh (\mu_1 - \mu_j)} B_{M-1}^{(j)} \right) \non
&=& - v^+(\mu_1) v^+(\l) B_M - \f 2{\sinh (\l - \mu _1)}
\left( e ^{-(\l - \mu_1)} X^+(\l) - e ^{\l - \mu_1} X^+(\mu_1) \right) B_M \non
&+& 2 X^+(\mu_1) \sum_{j=2}^{M+1} (-1)^j \f {e ^{-(\mu_1 - \mu_j)}}
{\sinh (\mu_1 - \mu_j)} v^+(\l) B_{M-1}^{(j)} \;.
\eaa
Now we can substitute the expressions for $v^+(\l) B_M$ and 
$v^+(\l) B_{M-1}^{(j)}$. After rearranging the terms in an
appropriate way we have
\baa \la{vBN+1b} 
v^+(\l) B_{M+1} &=& (-1)^{M+1} B_{M+1} v^+(\l) + 2 \sum_{j=1}^{M+1} 
\f {(-1)^j}{\sinh (\l - \mu _j)} \non
&&\times \left( e^{-(\l - \mu _j)} X^+(\l) - e^{-(\mu _j - \l)} 
X^+(\mu _j) \right) B_M^{(j)} \;. \non
\eaa
This completes the proof of the lemma.
{\qed}

The proofs of the other lemmas are analogous to the proofs we have
illustrated above. They do not contain illuminating insights and are 
considerably longer than the two we have seen. Thus, we will omit them.

The recurrence relation \Ref{recr} can be solved explicitly. 
To be able to express the solution in a compact form 
it is useful to introduce a contraction operator $d$.

\begin{definition} Let $d$ be a contraction operator whose 
action on an ordered product 
$\overset{M}{\underset{\longrightarrow}
{\underset{j = 1}{\prod}}} v^+(\mu_j)$, $M \geq 2$, 
is given by 
\begin{equation} \la{d}  
d \left(v^+(\mu_1) v^+(\mu_2) \dots  v^+(\mu_M) \right) = 
2 \sum_{j = 1}^{M-1} X^+(\mu_j) {\sum_{k=j+1}^{M}} (-1)^{\sigma(jk)}
\f{e^{-(\mu_j-\mu_k)}}{\sinh (\mu_j - \mu_k)} 
\underset{\longrightarrow}{\prod_{m \neq j, k}^{M}} 
v^+(\mu_m) 
\end{equation}
where $\s(jk) $ is the parity of the permutation 
$$ 
\s : (1, 2, \dots , j, j+1,  \dots , k, \dots , M) \to 
(1, 2, \dots , j, k, j+1, \dots , M) \;.
$$ 
\end{definition}
The $d$ operator can be applied on an ordered product
$\overset{M}{\underset{\longrightarrow}{\underset{j = 1}
{\prod}}}v^+(\mu_j)$ consecutively several times, up to 
$[M/2]$, the integer part of $M/2$. 
\begin{theorem}
Explicit solution to the recurrence relation \Ref{recr} is 
given by
\begin{equation} \la{srr}  
B_{M}(\mu_1, \dots , \mu_M) = 
\underset{\longrightarrow}{\prod_{j=1}^{M}}
v^+(\mu_j) + \sum_{m = 1}^{[M/2]} \f{1}{m!} \; d^m
\underset{\longrightarrow}{\prod_{j = 1}^{M}} v^+(\mu_j) 
= \exp d \; \underset{\longrightarrow}{\prod_{j=1}^{M}}
v^+(\mu_j)
\;.  
\end{equation}
\end{theorem}

The properties of the creation operators $B_M$ studied in the this Section 
will be fundamental tools in determining characteristics of the trigonometric 
$osp(1|2)$ Gaudin model. Our primary interest is to obtain the spectrum and 
the eigenvectors of the generating function of integrals of motion $t(\l)$ 
\Ref{tG}.


\section{Spectrum and eigenstates of trigonometric $osp(1|2)$ Gaudin model}

With the help of the creation operators $B_M$ it is possible to obtain
the eigenvectors as well as the corresponding eigenvalues of the Hamiltonians
of the trigonometric Gaudin model. This result is a direct consequence of 
the following theorem.

\begin{theorem}
The lowest spin vector $\Om _-$ \Ref{vac} is an eigenvector
of the generating function of integrals of motion $t(\l)$ 
\Ref{tG} with the corresponding eigenvalue $\L _0 (\l)$
\begin{equation}
\la{l0-t} 
t(\l) \, \Om _- = \L _0 (\l) \, \Om _- \;, \quad \L _0 (\l)= 
\rho^2(\l) + \rho'(\l) \;.   
\end{equation}
Furthermore, the action of the $B$-operators \Ref{recr} on the lowest 
spin vector $\Om _-$ yields the eigenvectors
\begin{equation} \la{eigv}  
\Psi (\mu_1, \dots , \mu_M) = B_{M}(\mu_1, \dots , \mu_M) 
\; \Omega _- \;, 
\end{equation}
of the $t(\l)$ operator 
\begin{equation} \la{eigeq}  
t(\l) \Psi (\mu_1, \dots , \mu_M) = 
\L (\l ; \, \{\mu_j\}_{j=1}^M) \, \Psi (\mu_1, \dots , \mu_M) 
\;, 
\end{equation}
with the eigenvalues
\begin{equation} \la{l-t} 
\L (\l ; \, \{\mu_j\}_{j=1}^M) = y ^2 + \partial _{\l} y \,,
\end{equation}
here 
\begin{equation} \la{y} 
y (\l ; \, \{\mu_j\}_{j=1}^M) = \rho (\l) + \sum _{k=1}^M \coth (\l - \mu_k) \,,
\end{equation}
provided that the Bethe equations are imposed on the parameters 
$\{\mu_j\}_{j=1}^M$ of the state \Ref{eigv}
\begin{equation} \la{B-eq} 
\b _M (\mu_j) = \rho (\mu_j) + \sum_{k \neq j}^M \coth (\mu_j - \mu_k) = 0 \;.   
\end{equation}
\end{theorem}
{\it Proof.} The equation \Ref{l0-t} can be checked by a 
direct substitution of the definitions of the operator 
$t(\l)$ and the lowest spin vector $\Om _-$, 
the equations \Ref{tG} and \Ref{vac}, respectively. 

To show the second part of the theorem, we use the equation 
\Ref{eigv} to express the Bethe vectors 
$\Psi (\mu_1, \dots , \mu_M)$
\begin{equation} \la{peeq}  
t(\l) \Psi (\mu_1, \dots , \mu_M) = t(\l) \; 
B_{M}(\mu_1, \dots , \mu_M) \; \Omega _- \;. 
\end{equation}
Our next step is to use the third property of the $B$-operators, 
the equation \Ref{tB}, and the definition of
the lowest spin vector $\Om _-$ the equation \Ref{vac} 
in order to calculate the action of the operator $t(\l)$ on the 
Bethe vectors when the Bethe equations \Ref{B-eq} are imposed 
\baa \la{preeq} 
t(\l)B_{M} \Om _- &=& B_{M} t(\l) \Om _-  \non
&+&  \left( 2 \rho (\l) \sum _{i=1}^M \coth (\l -\mu _i) 
+ 2 \sum_{i < j}^M \coth (\l -\mu _i)
\coth (\l -\mu _j) + M \right)  B_{M} \Om _- \,. \non
\eaa
We can express the first term on the right hand side since 
we know how the operator $t(\l)$ acts on the vector $\Om _-$, 
the equation \Ref{l0-t}. Thus we have
\be \la{proeeq} 
t(\l) \; B_{M} \Om _- = \L (\l ; \, \{\mu_j\}_{j=1}^M) \; B_{M} \Om _- \;,\\
\ee
with
\ben
\L (\l ; \, \{\mu_j\}_{j=1}^M) = \L _0 (\l) 
+ 2 \rho (\l) \sum _{i=1}^M \coth (\l -\mu _i) 
+ 2 \sum_{i < j}^M \coth (\l -\mu _i) \coth (\l -\mu _j) + M \;,
\een
and we complete the proof by expressing the eigenvalue as 
\ben
\L (\l ; \, \{\mu_j\}_{j=1}^M) =  y ^2 + \partial _{\l} y \;, 
\quad \hbox{with} \quad
y (\l ; \, \{\mu_j\}_{j=1}^M) = \rho (\l) + \sum _{k=1}^M \coth (\l - \mu_k) \;. 
\een
{\qed}

\begin{corollary}
In the trigonometric Gaudin realization given by the equations \Ref{Gr-h}, \Ref{Gr-v}, 
\Ref{Gr-X} and \Ref{Gre} the Bethe vectors $\Psi (\mu_1, \dots , \mu_M)$ 
\Ref{eigv} are the eigenvectors of the Gaudin Hamiltonians \Ref{sGh} (see also \c{LSU})
\baa \la{Geeq}  
H^{(a)} \Psi (\mu_1, \dots , \mu_M) = E^{(a)}_M \Psi (\mu_1, \dots , \mu_M) \;, 
\eaa
with the eigenvalues
\baa \la{GEn}
E ^{(a)}_M = \underset{b\neq a}{\sum_{b=1}^{N}} l_a \, l_b 
\coth (z _a - z _b) + \sum _{j=1}^M l_a \coth (\mu _j - z _a) \;,
\eaa 
when the Bethe equations are imposed
\baa \la{BeqG} 
\b _M (\mu_j) = \rho (\mu_j) + \sum_{k \neq j}^M \coth (\mu_j - \mu_k) = 
\sum_{a=1}^N (-l_a) \coth (\mu_j - z_a) + 
\sum_{k \neq j}^M \coth (\mu_j - \mu_k) = 0 \;. \non 
\eaa
\end{corollary}
{\it Proof.} The statement of the corollary follows from residue of the
equation \Ref{eigeq} at the point $\l = z _a$. The residue can be
determined using \Ref{tpole}, \Ref{l-t} and \Ref{l0-t}.
{\qed}

Comparing the eigenvalues $E^{(a)}_M$ \Ref{GEn} of the Gaudin Hamiltonians
and the Bethe equations \Ref{BeqG} with the corresponding quantities of
the $sl(2)$ Gaudin model \c{GB,S87} we arrive to an interesting observation.

\begin{remark}
The spectrum of the $osp(1|2)$ trigonometric Gaudin model with the spins 
$l_a$ coincides with the spectrum of the $sl(2)$ trigonometric Gaudin system 
for the integer spins (see an analogous observation for partition functions 
of corresponding anisotropic vertex models in \c{HS}). 
\end{remark}

\begin{remark}
The Bethe vectors are eigenstates of the global generator  $h _{gl}$
\be \la{hglee}
h _{gl} \Psi (\mu_1, \dots , \mu_M) = 
\left( - \sum _{a=1}^N l_a  + M \right) \Psi (\mu_1, \dots , \mu_M)
\;.
\ee
\end{remark}
As oppose to the $osp(1|2)$-invariant model \c{KM}, these Bethe vectors are 
not the lowest spin vectors of the global $osp(1|2)$ since they are 
not annihilated by the generator $v^-_{gl}$
\be \la{vglBv} 
v^-_{gl} \Psi (\mu_1, \dots , \mu_M) \neq 0 \;, 
\ee
once the Bethe equations are imposed \Ref{BeqG}. These conclusions follow
from the remark 3.1, in particular the equations \Ref{hglB} and \Ref{v-glB}, 
and the definition of the Bethe vectors \Ref{eigv}.  


As was pointed out already in \c{GB} for the $sl(2)$ case,
there are several modifications of the Hamiltonians (\ref{sGh}). One of them
is the Richardson's pairing-force Hamiltonian \c{Rich1, Rich2, GB}. 
These modifications can be formulated in the framework of the universal
$L$-operator and $r$-matrix formalism \Ref{rL} \c{S87}.

Due to invariance of the $r$-matrix \Ref{r} with respect to the Cartan element
\baa \la{r-inv}
\left[ r (\l) , h \otimes I + I \otimes h \right] = 0 \, , 
\quad h \in osp (1|2)
\eaa
one can add to the $L$-operator the element $h$
\baa \la{L-mod}
L (\l) \to \tilde{L} (\l) = g \, h +  L (\l) \, ,
\eaa
preserving commutation relations \Ref{rL}. Then 
\baa \la{t-mod}
\wtt (\l) = \half \, {\str} \, \widetilde{L} ^2(\l) = t (\l) 
+ 2 g \, h (\l) + g ^2\, ,
\eaa
will have the commutativity property, {\ie} $\wtt (\l) \wtt (\mu)
= \wtt (\mu) \wtt (\l)$. Hence we can take $\widetilde{t} (\l)$
to be the generating function of the (modified) integrals of motion
\baa \la{tpole-mod}
\wtt (\l) = ( h_{gl} - g ) ^2 + \sum_{a =1}^N \left( \f {c_2(a)}{\sinh ^2 (\l - z_a)}
+ \, 2 \, \f {e ^{\l -z_a}}{\sinh (\l - z_a)}  \, \widetilde{H}^{(a)} \right) \, , \\ 
\eaa
\be
\la{h-mod}
\widetilde{H}^{(a)} = \underset {\l=z_a} {\res} \wtt (\l) = 
g \; h _a +  H^{(a)} \,.  
\ee
In this case the eigenstates $\Psi _M$ are generated by the same
B-operators. However, corresponding eigenvalues and Bethe equations
are now given by
\be \la{l-t-mod}
\widetilde{\L} (\l ; \, \{\mu_j\}_{j=1}^M) = ( y + g ) ^2 + \p _{\l} y \, ,
\ee
here as before $y (\l ; \, \{\mu_j\}_{j=1}^M) = \sum_{a=1}^N {(-l_a)} 
\coth (\l - z_a) + \sum _{k=1}^M \coth (\l - \mu_k)$,
\be
\la{evGh-mod}
\widetilde{E} ^{(a)}_M = E ^{(a)}_M + g \, ( -l _a) \, ,
\ee
\be
\la{Beq-mod}
\sum_{a=1}^N {(-l_a)} \coth (\mu_j - z_a) + \sum_{k \neq j}^M 
\coth (\mu_j - \mu_k) + g = 0 \,.
\ee 
The crucial step in the proof of these equations is the observation
that the commutation relations between the operator $\wtt (\l)$ \Ref{t-mod}
and the creation operators $B_M$ are equal to the commutation relations 
\Ref{tB} but with modified operator $\hb _M(\mu_j) \to \hb _M(\mu_j) + g$.
To see this notice the similarity between the terms with $v ^+(\l) B ^{(i)}_{M-1}$ 
operators and with $X ^+(\l) B ^{(i,j)}_{M-2}$ operators 
in the lemma 3.2 the equation \Ref{hB} and in the lemma 3.3 the equation \Ref{tB}.

Richardson like Hamiltonian \c{Rich1, Rich2, GB, ALO, DS} can 
be obtained as a coefficient in the $\l \to + \infty$ expansion \c{KM}
\baa \la{t-mod-inf+}
\wtt (\l) &=& \left( h _{gl} + g \right) ^2 + 4 e ^{-2\l} \left( \left( h _{gl} - 1 + g \right) 
\left(\sum _a e ^{2z_a} h _a \right) \right. \non
&+& \left. 4 X ^+_{gl} \left(\sum _a e ^{2z_a} X ^-_a \right) + 
2 v ^+_{gl} \left(\sum _a e ^{2z_a}  v ^-_a\right) \right) + O \left(e ^{-4\l}\right) \, .
\eaa
Let us denote the coefficient next to the factor $4 e ^{-2\l}$ by $H _+$
\be \la{H+}
H _+ = \left( h _{gl} - 1 + g \right) \left(\sum _i e ^{2z_a} h _a \right)
+ 4 X ^+_{gl} \left(\sum _a e ^{2z_a} X ^-_a \right) + 
2 v ^+_{gl} \left( \sum _a e ^{2z_i}  v ^-_a \right) \, .
\ee
This Hamiltonian is obviously not symmetric. Similar Hamiltonian can be obtained 
as a coefficient in the $\l \to - \infty$ expansion
\baa \la{t-mod-inf-}
\wtt (\l) &=& \left( h _{gl} - g \right) ^2 + 4 e ^{2\l} \left( \left(\sum _a e ^{-2z_a} h _a \right) 
\left( h _{gl} - 1 - g \right) \right. \non
&+& \left. 4 \left(\sum _a e ^{-2z_a} X ^+_a \right) X ^-_{gl}  + 
2 \left(\sum _a e ^{-2z_a}  v ^+_a \right)  v ^-_{gl} \right) + O \left(e ^{4\l}\right) \, .
\eaa
Let us denote the coefficient next to the factor $4 e ^{2\l}$ by $H _-$, which
is also not symmetric. Thus, we choose the following symmetric combination
for a trigonometric generalization of the Richardson Hamiltonian
\baa \la{HR}
H _R &=& \f 12 \left( H _+ + H _- \right) = \left( h _{gl} - 1 \right) 
\left(\sum _a \cosh (2z_a) \; h _a \right) + g \; \left(\sum _a \sinh (2z_a) \; h _a \right) \non
&+& 2 \left(X ^+_{gl} \left(\sum _a e ^{2z_a} X ^-_a \right) +
\left(\sum _a e ^{-2z_a} X ^+_a \right) X ^-_{gl} \right) \non
&+& v ^+_{gl} \left(\sum _i e ^{2z_a}  v ^-_a \right) + 
\left(\sum _a e ^{-2z_a}  v ^+_a \right)  v ^-_{gl} \,.
\eaa
The eigenvalues of $H _R$ have different dependence on the quasi-momenta
from the rational case \c{Rich1,KM}
\be \la{HR-ev}
H _R \Psi_M(\mu_1, \ldots, \mu_M) = E_R (M) \Psi_M(\mu_1, \ldots, \mu_M) \,,
\ee
with
\baa
E_R (M) &=& \left( \sum _{j=1}^M \cosh ( 2 \mu_j ) - \sum_{a=1}^N l_a \cosh ( 2 z_a ) 
\right) \left( M - \sum_{a=1}^N l_a - 1 \right) \non
&+& \left( \sum _{j=1}^M \sinh ( 2 \mu_j ) 
- \sum_{a=1}^N l_a \sinh ( 2 z_a ) \right) g \,.
\eaa

More complicated modifications of Gaudin models can be obtained 
considering quasi - classical limit of the quantum spin system with 
non-periodic boundary conditions and corresponding reflection equation
\c{S88,KS91}. The $L$-operator can be expressed in terms of the original 
one \Ref{L} as
\be \la{L-open}
L ^{(bGM)}( \l ; \{ z _j\} ) = L ( \l ; \{ z _j\} ) - L ( - \l ; \{ z _j\} ) \;,
\ee
in the case of the open chain, and it will satisfy more complicated liner 
brackets, defining a subalgebra of the loop algebra \Ref{rL} (see also 
\c{Gould} and references therein).

Most of the trigonometric Gaudin model relations have their
counterparts in the rational $osp (1|2)$-invariant case. To show this
one takes a scaling limit $\l \to \e \l$, $z_a \to \e z_a$,
\be \la{scale}
\lim _{\e \to 0} \e L_{\mathrm {trig}} ( \e \l ; \{\e z_a\} ) 
= L_{\mathrm {inv}} ( \l ; \{ z_a\} ) \;,
\ee
and in this way one reproduces known results for the $osp (1|2)$-invariant
model. However, as we shall see in the next Section some relations 
of the invariant GM have quite complicated analogs in the trigonometric 
case (a generalization of KZ system to include a ``magnetic field'' 
parameter $g$, requires a difference dynamical equation \c{TV2,EV}).
Also the modified $L$-operator \Ref{L-mod} requires to scale the
parameter $g \to g/  \e$. 

Another modification can be obtained by performing the similarity 
transformation on the $r$-matrix \Ref{crm} by the tensor square 
of the element $\exp(t X ^+)$. Then the sacling limit $\l \to \e \l$, 
$t \to \xi / 2 \e $ results in a modified $r$-matrix
\be \la{mrm} 
{\hat r} ( \l ) = \f{c ^{\otimes}_2}{\l} + \xi 
\left(h \otimes X^+ - X^+ \otimes h - v^+ \otimes v^+ \right) \;.
\ee
The loop superalgebra will be modified, as well as corresponding Hamiltonians 
\Ref{Gh}. Similarly, the algebraic Bethe Ansatz will require changes, 
although the Bethe equations and the spectrum will be the same as in the 
$osp(1|2)$-invariant case (see the $sl(2)$ case in \c{K02}).

The expression of the eigenvectors of a solvable model in terms of 
local variables parameterized by sites of the chain or by space 
coordinates, is known as coordinate Bethe Ansatz \c{GB}. The coordinate 
representation of the Bethe vectors gives explicitly analytical 
dependence on the parameters $\{ \mu _i\}_1^M$ and $\{ z_a\}_1^N$
useful in a relation to the  Knizhnik-Zamolodchikov equation (Section 5).
Using the Gaudin realization \Ref{Gr-h}, \Ref{Gr-v}, \Ref{Gr-X} of the generators 
\ben
v ^+(\mu) = \sum_{a=1}^N \f {e^{\mu - z_a}}{\sinh (\mu - z_a)} \, v^+_a
\;, \quad
X ^+(\mu) = \sum_{a=1}^N \f {e^{\mu - z_a}}{\sinh (\mu - z_a)}  \, X ^+_a 
\;,  
\een
and the definition of the creation operators \Ref{srr},
one can get the coordinate representation of the $B$-operators: 
\be \la{coorB}  
B_{M}(\mu_1, \mu_2, ..., \mu_M) = \sum_{\pi} \left(v^+_{a_1}
\cdots v^+_{a_M} \right) _{\pi} \prod_{a=1}^N 
\vph (\{ \mu_m^{(a)}\}^{\mid \Ka _a\mid}_1 ; z_a ) \;,  
\ee
where the first sum is taken over ordered partitions $\pi$ 
of the set $(1, 2, \ldots , M)$ into subsets $\Ka _a$, 
$a= 1, 2, \ldots , N$, including empty subsets with the constraints
\ben
\bigcup_a \Ka _a = (1, 2, \ldots , M) \;, \quad 
\Ka _a \bigcap \Ka _b = \emptyset \quad \hbox{for} \; a \neq b \;.
\een
The corresponding subset of quasimomenta
\ben
\left( \mu _1^{(a)} = \mu _{j_1} , \mu _2^{(a)} = \mu _{j_2} ,
\ldots \mu _{\mid \Ka _a \mid}^{(a)} = \mu _{j_{\mid \Ka _a\mid }} 
; j _m \in \Ka _a \right) \;,
\een
where $\mid \Ka _a\mid$ is the cardinality of the subset $\Ka _a$,
and $j_k<j_{k+1}$, entering into the coordinate wave function
\ben
\vph (\{ \nu _m \}_1^{\mid \Ka \mid} ; z) = 
\sum _{\s \in {\cal S}_{\mid \Ka \mid}} (-1)^{p(\s)}
\f {e^{\nu_{\s(1)} - \nu_{\s(2)}}} {\sinh (\nu_{\s(1)} - \nu_{\s(2)})}
\f {e^{\nu_{\s(2)} - \nu_{\s(3)}}} {\sinh (\nu_{\s(2)} - \nu_{\s(3)})}
\cdots
\f {e^{\nu_{\s (\mid \Ka \mid)} - z}} {\sinh (\nu_{\s (\mid \Ka \mid)} - z)}
\;.  
\een
Due to the alternative sum over permutations $\s \in {\cal S}_{\mid \Ka \mid}$ 
this function is antisymmetric with respect to the quasi-momenta. Finally the 
first factor in \Ref{coorB}
\ben
\left(v^+_{a_1} \cdots v^+_{a_M} \right) _{\pi}
\een
means that for $j_m \in \Ka _a$ corresponding indices of $v^+_{a_{j_m}}$
are equal to $a$ so that  $v^+_{a_{j_m}} = v^+_a$. One can collect these
operators into product $\prod _{a=1}^N \left(  v^+_a \right) ^{\mid \Ka _a \mid}$,
consequently we have an extra sign factor $(-1) ^{p(\pi)}$.

This coordinate representation is similar to the representations 
obtained in \c{BF,FFR,ReshVar} for the Gaudin models related to 
the simple Lie algebras (see also \c{M97}). The $Z_2$-grading 
of superalgebra results in extra signs, while the complicated structure 
of the $B_M$-operators (for the $sl(2)$ Gaudin model they are just products 
of $B_1$-operators $B_1(\mu_j) = X ^+(\mu_j)$) is connected with 
the fact that $(v^+_j)^2 = X_j^+ \neq 0$ while for $j \neq k$ $v^+_j$ 
and $v^+_k$ anticommute. 


\section{Solutions to the Knizhnik-Zamolodchikov equation}

Correlation functions $\psi (z _1 , \dots , z_N)$ of a two dimensional
conformal field theory satisfy the Knizhnik-Zamolodchikov equation 
\c{KZ}
\baa \la{KZ}
\ka \; \p _{z_a} \; \psi (z _1 , \dots , z_N) &=& H ^{(a)}
\psi (z _1 , \dots , z_N) \;,
\eaa
where $H ^{(a)} (a = 1 , \dots , N)$ are the Gaudin Hamiltonians 
\Ref{sGh} and $\psi (z _1 , \dots , z_N)$
is a function of $N$ complex variables with its values in a tensor product
$\mathcal{H} = \underset {a=1}{\overset {N}{\otimes}} V_a^{(l_a)}$.

A relation between the Bethe vectors of the Gaudin model related to 
simple Lie algebras and the solutions to the
Knizhnik-Zamolodchikov equation is well known for
sometime \c{BF, FFR}. Approach used here to obtain solutions to the
Knizhnik-Zamolodchikov equation corresponding to conformal field 
theory and  Lie superalgebra $osp(1|2)$ starting from B-vectors \Ref{eigv} 
is based on \c{BF}.

A solution in question is represented as a contour integral over the variables
$\mu_1 , \dots , \mu_M$
\baa \la{KZ-sol}
\psi ( z_1, \dots z_N ) &=& \oint \dots \oint \; \ph ( \vm | \vz ) \Psi ( \vm | \vz )
\; d\mu _1 \dots d\mu _M \; , 
\eaa
where an integrating factor $\ph ( \vm | \vz )$ is a scalar function
\baa \la{factor}
\ph ( \vm | \vz ) = \prod _{i< j}^M \sinh ( \mu _i - \mu _j ) ^{\f 1{\ka}}
\prod _{a< b}^N \sinh  ( z _a - z _b ) ^{\f {l_al_b}{\ka}}
\left( \prod _{k=1}^M \prod _{c=1}^N \sinh ( \mu _k - z _c ) ^{\f {-l_c}{\ka}}
\right) \;, 
\eaa
and $\Psi ( \vm | \vz )$ is a Bethe vector \Ref{eigv} where the corresponding
Bethe equations are not imposed. 
 
As a first step in the proof that $\psi ( z_1, \dots z_N )$ given by \Ref{KZ-sol}
is a solution of \Ref{KZ} we differentiate the product $\ph \Psi$ with 
respect to $z_a$ and obtain
\baa \la{calc}
\p _{z_a } \left( \ph  \Psi \right) = \p _{z_a } \left( \ph  \right ) \Psi +
\ph  \p _{z_a } \left( \Psi \right ) \;.
\eaa
Using \Ref{factor} the first term on the right hand side can be calculated explicitly 
\baa
\ka \p _{z_a } \ph = \left(\underset{b\neq a}{\sum _{b=1}^N} l_a \, l_b 
\coth (z _a-z _b) - \sum _{j=1}^M l_a \coth (z _a - \mu _j) \right) \ph 
= E ^{(a)}_M \ph \;.
\eaa
Furthermore, taking a residue of \Ref{tB} at $\l = z_a$ we have
\baa
H ^{(a)} \Psi =  E ^{(a)}_M \Psi + \sum _{j=1}^M (-1) ^j 
\f {e ^{-(z_a-\mu _j)}}{\sinh (z_a-\mu _j)}
\b _M ( \mu _j ) {\tPsi} ^{(j,a)} \;,
\eaa
where
\baa
{\tPsi} ^{(j,a)} = \left( v ^+_a B ^{(j)}_{M-1} + 2 X ^+_a \sum_{k\neq j}^M 
(-1)^{k + \Th(j-k)}
\f{e ^{-(\mu _j - \mu _k)}}{\sinh (\mu _j - \mu _k)} B ^{(j,k)}_{M-2} \right) \; \Om _- \;. 
\eaa
Hence \Ref{calc} can be written as
\baa \la{der0}
\ka \p _{z_a } \left( \ph  \Psi \right) = H ^{(a)} \left( \ph  \Psi \right) +
\ph \sum _{j=1}^M (-1) ^j \f {e ^{\mu _j - z_a}}{\sinh (\mu _j - z_a)}
\b _M ( \mu _j ) {\tPsi} ^{(j,a)} + \ka \ph  \p _{z_a } \left( \Psi \right )\;.
\eaa
Moreover, from \Ref{factor} we also have
\baa \la{der1}
\ka \p _{\mu_j } \ph = \left(\sum _{a=1} ^N  (-l_a) \coth (\mu _j - z _a) 
+  \mathop{\sum _{k=1}} \limits_{j\neq k}\limits^M  \coth ( \mu_j - \mu _k) 
\right) \ph = \b  _M (\mu _j) \ph \;,
\eaa
and from the lemma 3.4 follows
\baa \la{der2}
\p _{z_a } \Psi = \sum _{j=1}^M (-1) ^{j} \p _{\mu_j } 
\left ( \f {e ^{\mu _j-z_a}}{\sinh (\mu _j - z_a)} {\tPsi} ^{(j,a)} \right )
\eaa
Thus, using \Ref{der1} and \Ref{der2}, we can combine the last two terms in  
\Ref{der0} into a sum of first oder derivatives in ${\mu_j }$ 
\baa \la{final}
\ka \p _{z_a } \left( \ph  \Psi \right) = H ^{(a)} \left( \ph  \Psi \right) +
\ka \sum _{j=1}^M (-1) ^{j} \p _{\mu_j } 
\left ( \f {e ^{\mu _j-z_a}}{\sinh (\mu _j - z_a)} \; \ph \, {\tPsi} ^{(j,a)} \right )\;.
\eaa
A closed contour integration of $\ph \Psi$ with respect to 
$\mu _1 , \dots ,\mu _M$
will cancel the contribution from the terms under the sum in \Ref{final}
and therefore $\psi ( z_1, \dots z_N )$ given by \Ref{KZ-sol} 
satisfies the Knizhnik-Zamolodchikov equation.

Conjugated Bethe vectors $(B_M \Om_-) ^{\ast}$ are entering into the
solution $\tpsi( z_1, \dots z_N )$ of the dual Knizhnik-Zamolodchikov equation
\baa \la{dKZ}
- \ka \f {\p}{\p z_a} \tpsi (z _1 , \dots , z_N) &=& 
\tpsi (z _1 , \dots , z_N) \; H ^{(a)} \;.
\eaa
The scalar product $\left( \tpsi (z _1 , \dots , z_N) \; , 
\; \psi (z _1 , \dots , z_N) \right)$ does not depend on $\{z_j\} _1^N$
and its quasi-classical limit $\ka \to 0$ gives the norm of the Bethe
vectors due to the fact that the stationary points of the contour integrals 
for $\ka \to 0$ are solutions to the Bethe equations \c{ReshVar}
\baa \la{KZBeq}
\f{\p S}{\p \mu_j} = \sum _{a=1} ^N  (-l_a) \coth (\mu _j - z _a) +  
\mathop{\sum _{k=1}} \limits_{j\neq k}\limits^M  \coth ( \mu_j - \mu _k) = 0 \;,
\eaa
\baa \la{S}
S ( \vm | \vz ) &=& \ka \ln \ph = \sum_{a < b}^N l_a l_b \ln \left( \sinh (z_a - z_b) \right) \non
&+& \sum _{i < j}^M \ln \left( \sinh ( \mu _i - \mu _j) \right) - \sum _{a=1}^N \sum _{j=1}^M
l _a \ln \left( \sinh (z_a - \mu _j) \right) \;. \non
\eaa
According to the remark in the end of Section 4 analytical properties of 
the Bethe vectors of the trigonometric $osp(1|2)$ Gaudin model coincide 
with the analytical properties of the trigonometric $sl(2)$ Gaudin model. 
Thus, the expression for the norm of the Bethe vectors $\Psi$ \Ref{eigv} 
obtained as the first term in the asymptotic expansion $\ka \to 0$ coincides 
also
\baa \la{norm}
&&\left( \Psi \; , \; \Psi \right) =  \det \left(
\f {\p ^2 S}{\p \mu_j \; \p \mu_k} \right) \;, \\
\f {\p ^2 S}{\p \mu_j^2} &=& \sum _{a=1}^N \f{l_a}{\sinh ^2 (\mu _j - z_a)} - 
\sum _{k \neq j}^M \f{1}{\sinh ^2 (\mu _j - \mu _k)} \;, 
\quad \f{\p ^2 S}{\p \mu_j \; \p \mu_k} = \f{1}{\sinh ^2 (\mu _j - \mu _k)} \;, \non
\eaa
for $j\neq k$.

Finally we notice that the modification of the Gaudin Hamiltonians we discussed
at the end of the previous Section, can be easily transfered to the corresponding
modification of the Knizhnik-Zamolodchikov equations. The modification \Ref{L-mod} 
for the $sl(2)$-invariant Gaudin model was studied in \c{BK} as a quantization of 
the Schlesinger system (see also \c{R92}). This modification is related with 
extra factor in the integrating scalar function \Ref{factor}
\be \la{f-i}
\ph _j = \exp \left(\f{S_j}{\ka}\right) \;, \quad j=0 , 1 \;,
\ee
where $S_0 = S$ \Ref{S} and
\be 
\la{S-1}
S _1 = S _0 + g \sum _{j=1}^M \mu _j - g \sum _{a=1}^N l_a z_a \;, \\
\ee
correspond to the modification \Ref{L-mod}. 

Moreover, following the lines of \c{BK}, one can try to extend the connection
between the KZ equation and the Guadin model based on the modified $L$-operator
\Ref{L-mod} by extending the KZ system to include an equation of the form
\be
\left ( \ka \f{\p}{\p g} - H _{Rich} \right) \psi = 0 \,.
\ee  
However, such a straightforward generalization has failed in the trigonometric 
case (see below). We can comment on the extension in the rational case \c{KM} 
as a scaling limit of the trigonometric Gaudin model \Ref{scale}. The 
equations of original KZ system are defined by mutually commuting differential 
operators (see \Ref{cross-casimir})
\be
\nabla _a = \ka \; \f{\p}{\p z_a} - \widetilde{H}^{(a)} = 
\ka \; \f{\p}{\p z_a} - g h_a - \sum _{b\neq a} 
\f {c^{\otimes}_2(a,b)}{z_a - z_b} \,.
\ee
The operator
\be \la{nabla-rich}
\nabla _g = \ka \; \f{\p}{\p g} - H _{Rich} = 
\ka \; \f{\p}{\p g} - \sum _{a=1}^N z_a h _a - \f 1{2g}
\left(c _2 (gl) - h _{gl} (h _{gl} -1) \right) \,,
\ee
is commuting pairwise with the operators $\nabla _a$. Thus in the
rational case the KZ system can be generalized to include the 
operator $\nabla _g$. 

To prove that the solution to the modified KZ system with $\phi _1$ given 
by \Ref{f-i} and \Ref{S-1} is a solution to the generalized KZ system
we have to extend the trigonometric KZ equations \Ref{KZ} with modified 
Hamiltonians $gh _a +  H^{(a)}$ along the lines of \c{TV2,EV}. 
A difference equation must be introduced
\be \la{d-ext}
K (z _1 , \dots , z_N; g) \; \psi ( z_1, \dots z_N ; g)
= \psi ( z_1, \dots z_N ; g - 2 \ka ) \;,
\ee
instead of \Ref{nabla-rich}.
The operator $K$ is defined on the space $\mathcal{H}$
\be \la{K-our}
K (z _1 , \dots , z_N; g) = \exp \left(- 2 \sum _{a=1}^N z_a h_a \right)
P( g ; h_{gl}, v^+_{gl} , v^-_{gl}) \;,
\ee
where the operator $P$ depends on the global generators of the subalgebra
$osp(1|2) \subset \mathcal{L}_t$, and is constructed form the extremal 
projector $p(h, v^+, v^-)$ by a shift of the Cartan generator (see Appendix).
We introduce only one $K$ operator since the rank of $osp(1|2)$ Lie 
superalgebra is one. In general case, of simple Lie superalgebra of 
rank $r$, one has to consider a set of $K _k$, $k = 1, \dots, r$
(see \c{TV2,EV}).  


\section{Conclusion} 

By analyzing the model related to the trigonometric $osp(1|2)$ classical 
$r$-matrix the algebraic Bethe Ansatz approach to the Gaudin models is reviewed. 
The results presented in this paper are is some sense analogous to the ones 
we obtained for the $osp(1|2)$-invariant model \c{KM}. In particular, 
a striking similarity between some of the most fundamental characteristics 
of this system and the $sl(2)$ trigonometric Gaudin model was confirmed. 
Although explicitly constructed creation operators $B_M$ \Ref{srr} of the 
Bethe vectors are complicated polynomials of the $L$-operator entries 
$v^+(\l)$ and $X^+(\l)$, the coordinate form of the eigenfunctions 
differs only in signs from the corresponding states in the case of $sl(2)$  
trigonometric model, being antisymmetric functions of the quais-momenta.
Moreover, the eigenvalues and the Bethe equations coincide, provided that 
the $sl(2)$ Gaudin model with integer spins is considered. Analogously, 
the KZ equations based on both trigonometric models and for the nontrivial 
magnetic field $g$ require extension of the system of equations by 
the dynamical difference equation. 

Let us point out that by the method presented in this paper one can 
construct explicitly creation operators of the Gaudin models 
related to trigonometric Izergin-Korepin $r$-matrix \c{KBI,T}
corresponding to the twisted affine algebra $A_2^{(2)}$. 
Similarly to the simple Lie algebra case solutions to the 
Knizhnik-Zamolodchikov equation were constructed from the Bethe vectors 
using algebraic properties of the creation operators $B_M$ and the 
Gaudin realization of the loop superalgebra ${\cal L}_t(osp(1|2))$. 
This interplay between the Gaudin model and the Knizhnik-Zamolodchikov
equation enabled us to determine the norm of eigenfunctions of the 
Gaudin Hamiltonians
\ben 
\parallel \Psi (\mu _1 , \dots \mu _M ; \{ z_a\}_1^N ) \parallel ^2 =
\det \left( \f{\p ^2 S}{\p \mu_j \; \p \mu_k} \right) \;.
\een
The difficult problem of correlation function calculation
for general Bethe vectors
\ben
{\cal C} \left( \{ \nu_j \}_1^M ; \{ \mu _i\}_1^M ; \{ \l_k\}_1^K \right) =
\left( \Om _- \; , \; B_M ^{\ast} (\nu _1 , \dots \nu _M ) \prod_{k=1}^K
h(\l_k) \, B_M (\mu _1 , \dots \mu _M ) \Om _- \right)
\een
was solved nicely for the $sl(2)$-invariant Gaudin model in \c{S99} using the
Gauss factorization of the loop algebra group element and the
appropriate Riemann-Hilbert problem. Although the corresponding 
factorization is known even for the quantum superalgebra $\mathcal {U}_q (osp(1|2))$ 
\c{DKS} the final expression of the correlation functions is
difficult to obtain due to the complicated structure of the creation
operators $B_M(\mu _1 , \dots \mu _M ) = \hbox{Poly} ( v ^+ , X^+)$ \Ref{srr}.
The study of this problem is in progress and the following expression 
for the scalar product of the Bethe states is conjectured 
(see \c{S99})
\ben
\left( \Om _- \; , \; B_M ^{\ast} (\nu _1 , \dots \nu _M ) 
B_M (\mu _1 , \dots \mu _M ) \Om _- \right) = \sum _{\s \in {\cal S}_M} 
(-1) ^{p(\s)} \det {\cal M}^{\s} \;,
\een
where the sum is over symmetric group ${\cal S}_M$ and $M\times M$ matrix
${\cal M}^{\s}$ is given by
\baan
{\cal M}^{\s}_{jj} &=& \f{e ^{\mu _j - \nu _{\s(j)}}} {\sinh (\mu _j - \nu _{\s(j)})} 
\bigl( \r(\mu_j) -\r( \nu _{\s(j)}) \bigr)
- \sum_{k\neq j}^M \f{e ^{\mu _j - \mu _k} \, e ^{-( \nu_{\s(j)} - \nu _{\s(k)})}} 
{\sinh (\mu _j - \mu _k) \, \sinh (\nu _{\s(j)} - \nu _{\s(k)})} \;, \\
{\cal M}^{\s}_{jk} &=& \f{e ^{\mu _j - \mu _k} \, e ^{-( \nu_{\s(j)} - \nu _{\s(k)})}} 
{\sinh (\mu _j - \mu _k) \, \sinh (\nu _{\s(j)} - \nu _{\s(k)})} \;,
\quad \hbox{for $j,k=1, 2, \ldots M$}.
\eaan

\section{Acknowledgements}

We acknowledge useful discussions and communications with N. ~Yu. ~Reshetikhin,
V. ~O. ~Tarasov and T. ~Takebe. This work was supported by the grant 
PRAXIS XXI/BCC/22204/99, INTAS grant N 99-01459 and FCT project POCTI/33858/MAT/2000.

\begin{appendix}

\mathon
\section{Appendix: Orthosymplectic Lie superalgebra $osp(1|2)$}
\mathoff
 
The rank of the orthosymplectic Lie algebra $osp(1|2)$ is one 
and its dimension is five \c{Ritt}. The three even generators are $h, X^+, X^-$ and 
the two odd generators are $v^+, v^-$. The (graded) commutation 
relations of the generators are 
\baa
\ba{cc} \la{osp}
\left[ h, X^{\pm} \right] _- = \pm 2 X^{\pm} \;, \quad 
\left[ X^+, X^- \right] _-  = \, h \;, \\ 
\left[ h, v^{\pm} \right] _- = \pm v^{\pm} \;, \quad    
\left[ v^+, v^- \right] _+ = - h \;, \\
\left[ X^{\mp}, v^{\pm} \right] _- = \, v^{\mp} \;, \quad    
\left[ v^{\pm}, v^{\pm} \right] _+ = {\pm} 2 X^{\pm} \;,  \\
\left[ X^{\pm}, v^{\pm} \right] _- = 0 \;, 
\ea
\eaa

The Casimir element is 
\baa \la{Cas}  
c_2 &=& h^2 + 2 \left(X^+ X^- + X^- X^+ \right) 
+ \left(v^+ v^- - v^- v^+ \right)  \nonumber\\
&=& h^2 - h + 4 X^+ X^- + 2 v^+ v^-  \,.
\eaa
It is interesting to point out the existance of a ``square root'' of this
element
\be \la{c1} 
c_1 = h + 2 v^+ v^- - \f{1}{2} \, , \quad 
\left( c_1 \right) ^2 = c _2 + \f{1}{4} \,,
\ee
with a grading property $[c_1 , X^{\pm}] = 0$,  $[c_1 , h] = 0$ and
$c _1  v^{\pm} = -  v^{\pm}c _1$.
The finite dimensional irreducible representations $V^{(l)}$ of the 
$osp(1|2)$ Lie superalgebra are parameterized by an integer $l$, so that their 
dimensions $2l + 1$ and the values of the Casimir element 
\Ref{Cas} $c_2 = l(l+1)$ coincide with the same characteristics 
of the integer spin $l$ irreducible representations of $sl(2)$.

The fundamental irreducible representation $V ^{(1)}$ of $osp(1|2)$ 
is three dimensional. We choose a grading 
of the basis vectors $e_j;\, j=1,2,3$ to be $(0, 1, 0)$.
Explicitly we have
\baan
h &=& \left( \ba{ccc}
1 & 0 & 0 \\
0 & 0 & 0 \\
0 & 0 & -1 \ea \right) \;, 
\\
v_-=(v_+) ^{st} &=& \left( \ba{ccc}
 0 & 0 & 0 \\
-1 & 0 & 0 \\
 0 & 1 & 0 \ea \right) \:,
\eaan
together with $X ^{\pm} = \pm (v ^{\pm})^2$. The matrix $v^+$ in the
representation $V ^{(l)}$ has $2l$ non-zero elements on the second upper 
diagonal only, and these elements are 
\be
\{ (v^+)_{jj+1}\} = \left(
\sqrt {l}, \sqrt {1}, \sqrt {l-1}, \sqrt {2},  \dots, \sqrt {1} , 
\sqrt {l} \right)\;, \quad j = 1, 2, \dots , 2l \;.
\ee

The extermal projector \c{BT} for $osp(1|2)$ (on the lowes weight vectors):
\baa \la{ex-proj}
p(h,v^+,v^-) &=& \sum _{k=0}^{\infty} \f {(-1)^k}{k!}
\left( (v^+)^{2k}(v^-)^{2k} +  (v^+)^{2k+1}(v^-)^{2k+1} 
\f 1{h-k-1} \right) \prod _{j=1}^k \f 1{h-j} \non
&=& \left( 1 + v^+ v^- \f 1{h-1} \right) \left( \sum_{k=0}^{\infty} 
(X^+)^{k}(X^-)^{k} \prod _{j=1}^k \f 1{j(h-j-1)} \right) \non
&=& p_s (h,v^+,v^-) p _0 (h,X^+,X^-) \;,
\eaa
here $p _0 (h,X^+,X^-)$ is the usual $sl(2)$ extermal projector.
There is no such factorization property for the extermal projector
of the quantum superalgebra $\mathcal{U}_q (osp(1|2))$ \c{KR}.
\end{appendix}

\end{document}